\documentclass[twocolumn]{aastex63}
\usepackage{graphicx}
\graphicspath{{figure/}}
\usepackage{subfigure}
\usepackage{color, hyperref, epsfig}
\usepackage{apjfonts, natbib}
\usepackage{appendix}
\usepackage{amsmath}
\usepackage{float}
\usepackage{bm}
\usepackage{booktabs}
\usepackage{multirow}
\usepackage{ragged2e}
\usepackage{lineno}

\newcommand{\lum}{erg\,s\ensuremath{^{-1}}}
\newcommand{\msun}{\ensuremath{M_{\odot}}}

\shorttitle{WFST TDEs}
\shortauthors{Wu et al.}

\begin{document}

\title{Two Faint Tidal Disruption Events Discovered in a Pilot Search with the Wide Field Survey Telescope}

\author[0009-0006-8112-0187]{Mingxin Wu}
\affiliation{Department of Astronomy, University of Science and Technology of China, Hefei, 230026, China; wmx112611226@mail.ustc.edu.cn, jnac@ustc.edu.cn}
\affiliation{School of Astronomy and Space Sciences,
University of Science and Technology of China, Hefei, 230026, China}

\author[0000-0002-7152-3621]{Ning Jiang}
\affiliation{Department of Astronomy, University of Science and Technology of China, Hefei, 230026, China; wmx112611226@mail.ustc.edu.cn, jnac@ustc.edu.cn}
\affiliation{School of Astronomy and Space Sciences,
University of Science and Technology of China, Hefei, 230026, China}

\author[0000-0003-3824-9496]{Jiazheng Zhu}
\affiliation{Department of Astronomy, University of Science and Technology of China, Hefei, 230026, China; wmx112611226@mail.ustc.edu.cn, jnac@ustc.edu.cn}
\affiliation{School of Astronomy and Space Sciences,
University of Science and Technology of China, Hefei, 230026, China}

\author[0000-0002-6221-1829]{Jianwei Lyu}
\affiliation{Steward Observatory, University of Arizona, 933 North Cherry Avenue, Tucson, AZ 85721, USA}

\author[0000-0003-4959-1625]{Zheyu Lin}
\affiliation{Université Paris Cité, CNRS, Astroparticule et Cosmologie, F-75013 Paris, France}
\affiliation{Department of Astronomy, University of Science and Technology of China, Hefei, 230026, China; wmx112611226@mail.ustc.edu.cn, jnac@ustc.edu.cn}
\affiliation{School of Astronomy and Space Sciences,
University of Science and Technology of China, Hefei, 230026, China}

\author[0000-0003-4225-5442]{Yibo Wang}
\affiliation{Department of Astronomy, University of Science and Technology of China, Hefei, 230026, China; wmx112611226@mail.ustc.edu.cn, jnac@ustc.edu.cn}
\affiliation{School of Astronomy and Space Sciences, University of Science and Technology of China, Hefei, 230026, China}

\author[0000-0002-1517-6792]{Tinggui Wang}
\affiliation{Department of Astronomy, University of Science and Technology of China, Hefei, 230026, China; wmx112611226@mail.ustc.edu.cn, jnac@ustc.edu.cn}
\affiliation{School of Astronomy and Space Sciences,
University of Science and Technology of China, Hefei, 230026, China}

\author[0000-0003-4700-348X]{Chengyi Wang}
\affiliation{Department of Astronomy, University of Science and Technology of China, Hefei, 230026, China; wmx112611226@mail.ustc.edu.cn, jnac@ustc.edu.cn}
\affiliation{School of Astronomy and Space Sciences,
University of Science and Technology of China, Hefei, 230026, China}

\author[0009-0003-9474-8457]{Weiyu Wu}
\affiliation{Department of Astronomy, University of Science and Technology of China, Hefei, 230026, China; wmx112611226@mail.ustc.edu.cn, jnac@ustc.edu.cn}
\affiliation{School of Astronomy and Space Sciences,
University of Science and Technology of China, Hefei, 230026, China}

\author[0000-0003-3965-6931]{Jie Lin}
\affiliation{Department of Astronomy, University of Science and Technology of China, Hefei, 230026, China; wmx112611226@mail.ustc.edu.cn, jnac@ustc.edu.cn}
\affiliation{School of Astronomy and Space Sciences, University of Science and Technology of China, Hefei, 230026, China}

\author{Zelin Xu}
\affiliation{Department of Astronomy, University of Science and Technology of China, Hefei, 230026, China; wmx112611226@mail.ustc.edu.cn, jnac@ustc.edu.cn}
\affiliation{School of Astronomy and Space Sciences,
University of Science and Technology of China, Hefei, 230026, China}

\author[0000-0003-4721-6477]{Minxuan~Cai}
\affiliation{Department of Astronomy, University of Science and Technology of China, Hefei, 230026, China; wmx112611226@mail.ustc.edu.cn, jnac@ustc.edu.cn}
\affiliation{School of Astronomy and Space Sciences,
University of Science and Technology of China, Hefei, 230026, China}

\author[0009-0006-4622-1417]{Ranfang Zheng}
\affiliation{Department of Astronomy, University of Science and Technology of China, Hefei, 230026, China; wmx112611226@mail.ustc.edu.cn, jnac@ustc.edu.cn}
\affiliation{School of Astronomy and Space Sciences,
University of Science and Technology of China, Hefei, 230026, China}

\author[0000-0003-4200-4432]{Lulu fan}
\affiliation{Department of Astronomy, University of Science and Technology of China, Hefei, 230026, China; wmx112611226@mail.ustc.edu.cn, jnac@ustc.edu.cn}
\affiliation{School of Astronomy and Space Sciences,
University of Science and Technology of China, Hefei, 230026, China}
\affiliation{Institute of Deep Space Sciences, Deep Space Exploration Laboratory, Hefei 230026, China}
\affiliation{College of Physics, Guizhou University, Guiyang 550025, China}

\author[0000-0002-9092-0593]{Ji-an Jiang}
\affiliation{Department of Astronomy, University of Science and Technology of China, Hefei, 230026, China; wmx112611226@mail.ustc.edu.cn, jnac@ustc.edu.cn}
\affiliation{School of Astronomy and Space Sciences,
University of Science and Technology of China, Hefei, 230026, China}
\affiliation{National Astronomical Observatory of Japan, National Institutes of Natural Sciences, Tokyo 181-8588, Japan}

\author[0000-0002-7660-2273]{Xu Kong}
\affiliation{Department of Astronomy, University of Science and Technology of China, Hefei, 230026, China; wmx112611226@mail.ustc.edu.cn, jnac@ustc.edu.cn}
\affiliation{School of Astronomy and Space Sciences,
University of Science and Technology of China, Hefei, 230026, China}
\affiliation{Institute of Deep Space Sciences, Deep Space Exploration Laboratory, Hefei 230026, China}

\author{Feng Li}
\affiliation{State Key Laboratory of Particle Detection and Electronics, University of Science and Technology of China, Hefei 230026, China}

\author{Ming Liang}
\affiliation{National Optical Astronomy Observatory (NSF’s National Optical-Infrared Astronomy Research Laboratory) 950 N Cherry Ave. Tucson Arizona 85726, USA}

\author[0000-0002-0561-7937]{Binyang Liu}
\affiliation{Purple Mountain Observatory, Chinese Academy of Sciences, Nanjing 210023, China}
\affiliation{School of Astronomy and Space Sciences,
University of Science and Technology of China, Hefei, 230026, China}

\author[0000-0002-3105-3821]{Zhen Wan}
\affiliation{Department of Astronomy, University of Science and Technology of China, Hefei, 230026, China; wmx112611226@mail.ustc.edu.cn, jnac@ustc.edu.cn}
\affiliation{School of Astronomy and Space Sciences,
University of Science and Technology of China, Hefei, 230026, China}

\author{Hairen Wang}
\affiliation{Purple Mountain Observatory, Chinese Academy of Sciences, Nanjing 210023, China}
\affiliation{School of Astronomy and Space Sciences, University of Science and Technology of China, Hefei, 230026, China}

\author[0000-0003-1617-2002]{Jian Wang}
\affiliation{Institute of Deep Space Sciences, Deep Space Exploration Laboratory, Hefei 230026, China}
\affiliation{State Key Laboratory of Particle Detection and Electronics, University of Science and Technology of China, Hefei 230026, China}

\author[0000-0002-1935-8104]{Yongquan Xue}
\affiliation{Department of Astronomy, University of Science and Technology of China, Hefei, 230026, China; wmx112611226@mail.ustc.edu.cn, jnac@ustc.edu.cn}
\affiliation{School of Astronomy and Space Sciences,
University of Science and Technology of China, Hefei, 230026, China}

\author[0000-0002-1463-9070]{Hongfei Zhang}
\affiliation{State Key Laboratory of Particle Detection and Electronics, University of Science and Technology of China, Hefei 230026, China}

\author[0000-0002-1330-2329]{Wen Zhao}
\affiliation{Department of Astronomy, University of Science and Technology of China, Hefei, 230026, China; wmx112611226@mail.ustc.edu.cn, jnac@ustc.edu.cn}
\affiliation{School of Astronomy and Space Sciences,
University of Science and Technology of China, Hefei, 230026, China}

\author[0000-0003-3728-9912]{Xianzhong Zheng}
\affiliation{Tsung-Dao Lee Institute and Key Laboratory for Particle Physics, Astrophysics and Cosmology, Ministry of Education, Shanghai Jiao Tong University, Shanghai, 201210, China}

\author[0000-0003-0694-8946]{Qingfeng Zhu}
\affiliation{Department of Astronomy, University of Science and Technology of China, Hefei, 230026, China; wmx112611226@mail.ustc.edu.cn, jnac@ustc.edu.cn}
\affiliation{School of Astronomy and Space Sciences,
University of Science and Technology of China, Hefei, 230026, China}


\begin{abstract}

We report the discovery of TDE~2025alrm/WFST250129blnve and TDE~2025amkg/WFST251021dbmwk in the first six months of the wide-field survey (WFS) performed by the 2.5-meter Wide Field Survey Telescope (WFST). Both events are apparently faint among known optical tidal disruption events (TDEs), with peak magnitudes of $g \sim 20.0$ and $g \sim 19.6$, respectively. 
Blackbody fits to the multi-band photometry yield roughly constant temperatures of $\sim 10^{4}$ K for both sources, with peak bolometric luminosities of $10^{43.6}$ erg s$^{-1}$ and $10^{43.8}$ erg s$^{-1}$, respectively. Optical spectra obtained during the declining phase display characteristic broad Balmer, He II, and N III emission lines, classifying both sources as H+He TDEs. Host-galaxy SED fitting suggests stellar masses on the order of $10^{9}$~\msun\ and black hole masses of $\sim10^{6}$~\msun, which are typical of optically selected TDE hosts. The pilot discovery of these two faint TDEs along with dozens of candidates, demonstrates the capability of WFST to discover TDEs, owing to its exceptional survey depth. Furthermore, these events highlight the value of the $u$ band in multi-band surveys for TDE selection, particularly for distinguishing TDEs from supernovae at early phases. With continued operations, WFST is expected to substantially expand the known TDE population, especially by uncovering nearby low-luminosity events and high-redshift TDEs, both of which remain poorly explored.

\end{abstract}

\keywords{High energy astrophysics (739); Tidal disruption (1696); Time domain astronomy (2109)}

\section{Introduction}

It is widely believed that supermassive black holes (SMBHs) reside at the centers of most galaxies. When a star approaches an SMBH closely enough, it can be tidally disrupted as the black hole’s tidal forces overcome the star’s self-gravity, an event known as a tidal disruption event (TDE)~\citep{Rees1988}. Part of the stellar debris subsequently falls back onto the black hole and forms an accretion disk. This process generates a luminous flare in the UV/optical and X-ray bands, providing a powerful way to detect quiescent black holes. Thanks to wide-field time-domain surveys over the past decade, the discovery and study of TDEs have advanced significantly~\citep{Gezari2021}. In particular, the Zwicky Transient Facility (ZTF; \citealt{Masci2019}) has dramatically increased the TDE discovery rate, enabling systematic studies of TDEs based on large samples (e.g., \citealt{vV2021, Hammerstein2023, Yao2023}). These systematic studies have revealed substantial spectroscopic diversity among TDEs. Their optical spectra are typically characterized by a blue continuum accompanied by broad hydrogen and/or helium emission lines, based on which they are commonly classified as TDE-H, TDE-H+He, or TDE-He. Some TDEs exhibit nearly featureless blue continua with weak or absent broad emission lines and are classified as TDE-featureless.

Optically selected TDEs typically exhibit a set of characteristic photometric properties that have become the basis for identifying TDE candidates in optical time-domain surveys. These events are generally characterized by blackbody temperatures of $(1$--$5)\times10^4\ \mathrm{K}$, resulting in persistently blue optical colors (e.g., $g-r<0$) with little temporal evolution. Most TDEs rise to peak brightness on timescales of about one month, reaching peak blackbody luminosities of $10^{43}$--$10^{46}$~\lum, followed by a power-law decline lasting months to years \citep{Lin2022b, Hammerstein2023, Yao2023}. However, the optical/ultraviolet spectral energy distributions (SEDs) of TDEs generally show no evidence of a turnover at the blue end \citep{Lin2025, Zhu2025}, suggesting that the SED peak likely resides in the extreme-ultraviolet (EUV) band. This is independently supported by infrared echo observations of several TDEs. The dust heating energy inferred from the infrared emission exceeds that directly observed in the optical/UV band, implying a substantial contribution from unobserved EUV radiation (e.g., \citealt{Wu2025}). Therefore, photometric observations in bluer bands, such as the $u$ band, are expected to be more effective for identifying TDEs than observations at redder wavelengths.

The currently known TDE population is largely confined to the low-redshift Universe ($\text{z}\lesssim0.5$) and is dominated by intrinsically luminous events (\,$\text{L}_\text{bb}\gtrsim10^{43}$~\lum\,) \citep{Yao2023}. Expanding the census toward fainter TDEs, either intrinsically underluminous or located at higher redshifts, will be a key focus of future TDE searches. With its large field of view of 6.5 $\text{deg}^2$, deep single-exposure limiting magnitude of $r\sim23$, and high-cadence multi-band imaging including the $u$ band over a wide area, the Wide Field Survey Telescope (WFST; \citealt{Wang2023}) is well suited to meet this requirement and to substantially expand the TDE sample toward both lower luminosities and higher redshifts \citep{Lin2022a}. 
Here we report the discovery of TDE 2025alrm (WFST250129blnve, $z=0.179$) and TDE 2025amkg (WFST251021dbmwk, $z=0.180$), two apparently faint TDEs identified by the Wide Field Survey (WFS) Program of WFST. Under this program, WFST will perform daily two-band observations: during dark or gray nights, it uses the $u$ band together with an additional optical band; during bright nights, it observes in the $g$ and $r$ bands. The survey covers approximately 3500 $\text{deg}^2$ with a cadence of 3–4 days and will reach a total surveyed area of 8000 $\text{deg}^2$ within one year. This observing strategy enables the discovery of large numbers of faint transients while providing sufficiently dense temporal sampling to robustly characterize their light-curve evolution, including both the rise and decline phases. With observed peak magnitudes of $g\sim20$ and $g\sim19.6$, TDE 2025alrm and TDE 2025amkg are among the faintest TDEs discovered to date in terms of apparent magnitude. Moreover, the well-sampled and complete optical light curves obtained by WFST demonstrate its capability to both detect and characterize faint TDEs, highlighting its power in probing the faint end of the TDE population.

In this work, we present our discovery, follow-up observations, and data analysis of TDE 2025alrm and TDE 2025amkg.
We assume a cosmology with $H_{0} =70$ km~s$^{-1}$~Mpc$^{-1}$, $\Omega_{m} = 0.3$, and $\Omega_{\Lambda} = 0.7$.

\section{Observations and Data}

\subsection{Selection of WFST TDE candidates}

To identify TDE candidates in the WFST database, we apply a set of selection criteria. Unlike previous studies that imposed a nuclear location requirement \citep{Hung2018, Vv2019}, we do not enforce such a constraint, as an increasing number of off-nuclear TDEs have been discovered in recent years \citep{Lin2018,Jin2025arXiv,Yao2025,Stein2026,Li2026}. Although the rise and decline timescales can help distinguish TDEs from SNe \citep{Vv2019}, we require only a clear and sustained rising or fading trend in the light curve. Specifically, we fit the photometric data before and after the maximum flux with a linear function and require the fitted slope to show the corresponding trend, with more than 70$\%$ adjacent measurements varying in the same direction. This criterion enables the identification of TDE candidates even when they are observed only during the early rising phase or the late declining phase. The colors of TDEs presented by Wu, W.Y. et al. (2026, in preparation) are distributed within the ranges of $-0.5<g-r<0$, $-0.3<u-g<0.2$, and $-0.7<u-r<0$. Since WFST generally conducts the WFS in two bands each night, one color measurement can be obtained for each epoch. To account for the effects of photometric uncertainties, host galaxy extinction and Galactic extinction, we adopt more conservative color cuts of $-0.6<g-r<0.1$, $-0.5<u-g<0.4$, and $-0.9<u-r<0.2$. A transient is considered to satisfy the color criterion if at least $80\%$ of its color measurements fall within the adopted color cuts. We further exclude sources with significant proper motions ($\mathrm{S/N}_{\rm PM}>5$ in Gaia) to reduce stellar contamination. To suppress contamination from AGNs, we require host infrared color to satisfy $\mathrm{W1}-\mathrm{W2}<0.5$. Based on these criteria, TDE candidates are selected using a database service originally developed for variable star studies in WFST observations~\citep{Linjie2025}, which also enables search and filtering of transient sources. Applying the selection criteria described above, we recovered 8 of the 11 spectroscopically confirmed TDEs identified by WFST during this six-month period and selected an additional 16 TDE candidates. These objects are discussed in Sections \ref{selection} and \ref{rate}.

\begin{figure*}
\centering
\begin{minipage}{1\textwidth}
\centering{\includegraphics[width=0.9\textwidth]{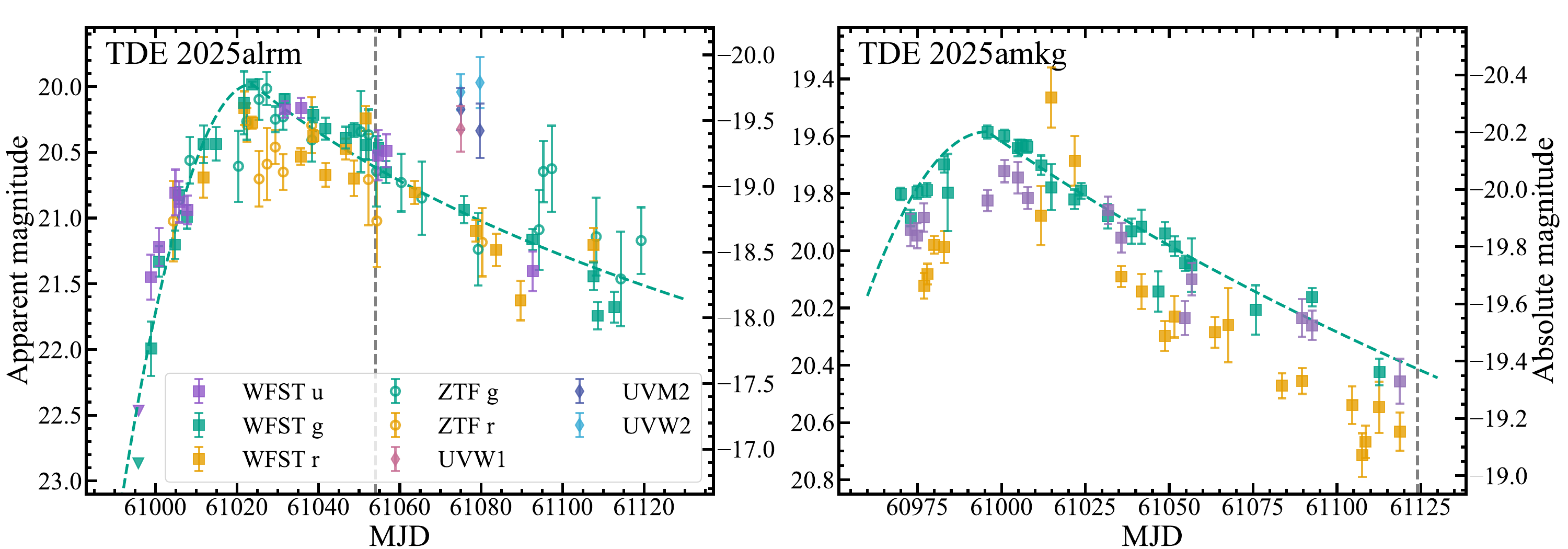}}
\end{minipage}
\caption{The multiwavelength light curves of TDE 2025alrm and TDE 2025amkg. The WFST, Swift/UVOT and ZTF photometry is denoted in different colors. All magnitudes are in the AB magnitude system. The gray vertical dashed lines indicate the observation time of MMT. The green dashed lines denote the best fit of WFST $g$ band light curves (see Equation \ref{lightcurve}).}
\label{ouvLC}
\end{figure*}

\subsection{WFST Discovery and Photometry}

TDE~2025alrm and TDE~2025amkg were first detected by WFST on 2025 November 19 and October 21, respectively, and were initially assigned the internal designations WFST250129blnve and WFST251021dbmwk in the WFST alert system. TDE~2025alrm is located at $\alpha = 08^{\mathrm h}13^{\mathrm m}00.649^{\mathrm s} \pm 0.137''$ and $\delta = +21^\circ09^\prime09.927^{\prime\prime} \pm 0.190''$ (J2000), with an angular offset of $0.074''\pm0.234''$ from the host galaxy center. Similarly, TDE~2025amkg is located at $\alpha = 08^{\mathrm h}20^{\mathrm m}10.118^{\mathrm s} \pm 0.163''$ and $\delta = +31^\circ24^\prime22.884^{\prime\prime} \pm 0.180''$, with an angular offset of $0.102''\pm0.243''$ from the host galaxy center. Both sources lie within the region of the WFST WFS, and their imaging data were obtained as part of the survey’s routine observations. 
The photometry was generated by the WFST data reduction pipeline and measured on the difference images~\citep{Cai2025}. Although the nominal cadence of the WFS is 3-4 days, the sky coverage of adjacent observations may sometimes overlap, leading to multiple observations of some sources within a single night. Therefore, data were further binned to a one-day cadence to maintain dense light-curve sampling while improving the S/N. We corrected all photometric data for Galactic extinction using the \citet{Cardelli1989} extinction law with $R_V = 3.1$ and $E(B-V) = 0.0402$ and $0.0401\ \text{mag}$, respectively \citep{Schlafly2011}. Both sources have relatively faint peak apparent magnitudes of $g \sim 20.0$ and $g \sim 19.6$, respectively (see Figure~\ref{ouvLC}). Benefiting from the high sensitivity and high cadence of WFST, their light curves are well sampled, demonstrating the capability of WFST to detect low-luminosity or high-redshift TDEs.

\subsection{ZTF Photometry}

We also collected host-subtracted light curves of TDE 2025alrm from ZTF. The ZTF $g$ and $r$ band light curves were obtained using the ZTF Forced Photometry Service \citep{Masci2023}. We binned data in one day intervals to improve the S/N and excluded measurements with $\mathrm{S/N} < 3$. No ZTF data are available for TDE~2025amkg (see Figure \ref{ouvLC}).

\subsection{Optical Spectroscopy}

\begin{figure*}
\centering
\begin{minipage}{1\textwidth}
\centering{\includegraphics[width=0.9\textwidth]{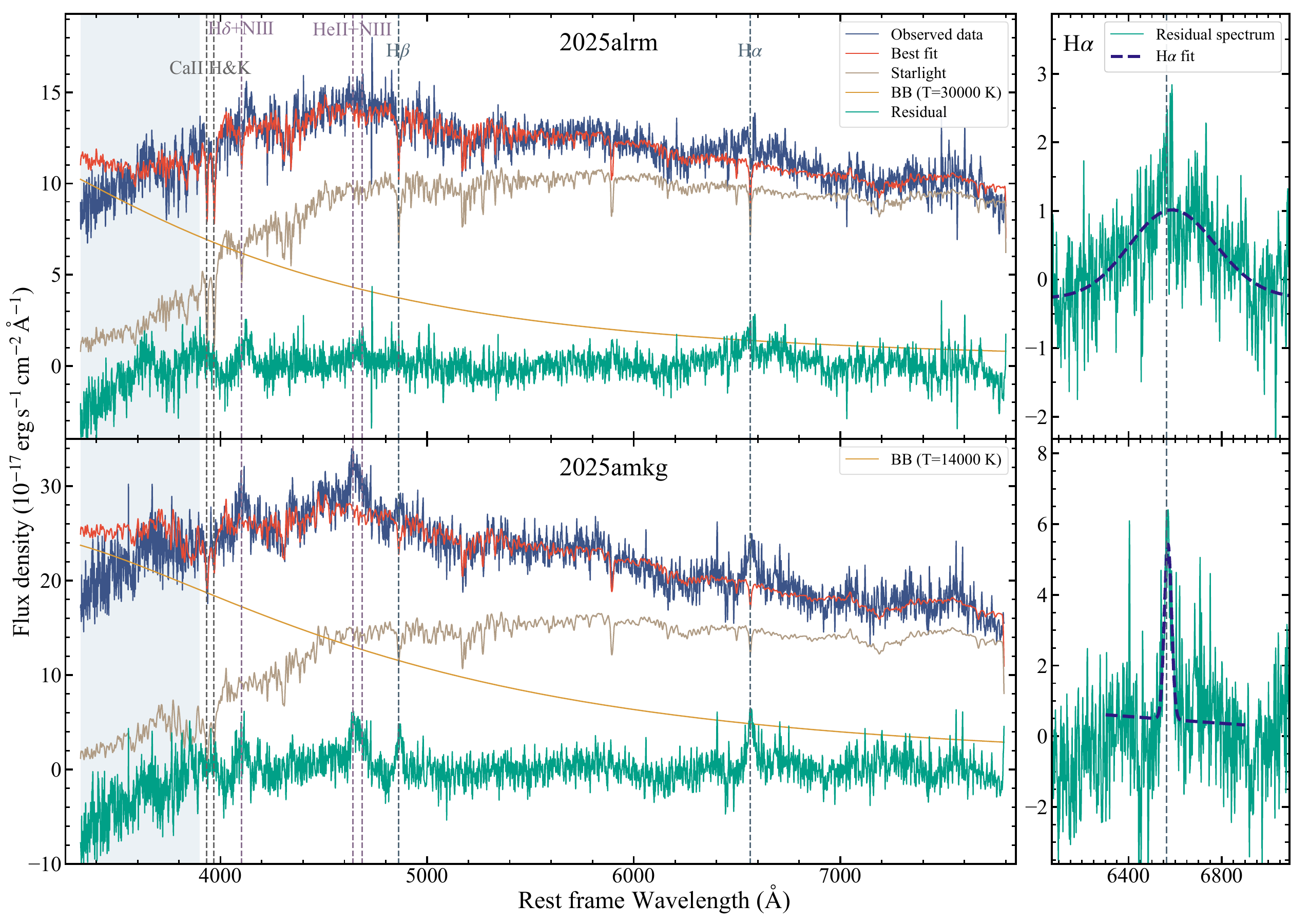}}
\end{minipage}
\caption{Left panel: The optical spectra of 2025alrm (top) and 2025amkg (bottom). The observed spectra are shown in blue, while the best-fitting models are shown in red. The stellar and blackbody components are shown in brown and yellow, respectively, and the continuum-subtracted residual spectra are shown in green. The vertical dashed lines mark the rest-frame wavelengths of prominent emission and absorption features. The shaded region indicates the wavelength range affected by poor flux calibration and excluded from the fitting. Right panel: Zoomed-in views of the H$\alpha$ emission lines in the residual spectra (green). The red dashed curves represent the best-fitting single-Gaussian models with a local baseline to account for residual continuum offsets. The gray dashed line marks the rest-frame wavelength of H$\alpha$.}
\label{MMT}
\end{figure*}

We triggered optical spectroscopic observations of TDE~2025alrm and TDE~2025amkg with the BINOSPEC spectrograph \citep{Fabricant2019} mounted on the 6.5m Multiple Mirror Telescope (MMT) on 2026 January 14 and March 25, respectively. Total exposure times of 3600~s and 2000~s were obtained for the two targets. For both observations, we used the 270 line mm$^{-1}$ grating ($R\sim1400$) centered at 6500~\AA\ together with a 1\arcsec\ slit. The data were reduced using the standard Binospec IDL pipeline maintained by the Smithsonian Astrophysical Observatory (SAO). Note that MMT/Binospec spectra are known to have flux-calibration issues at the blue end ($\lesssim4600$ \AA \ in the observed frame), which can make the continuum appear artificially redder. We therefore excluded this region from our analysis. We corrected the telluric absorption features, particularly the $\text{O}_{2}$ A and B bands using TelFit \citep{Gullikson2014}. The resulting spectra are presented in Figure~\ref{MMT}.

\subsection{Swift/UVOT}

Following the spectroscopic classification of TDE 2025alrm, we promptly triggered ultraviolet (UV) and X-ray follow-up observations with the \emph{Neil Gehrels Swift Observatory} (hereafter \emph{Swift}). However, only two epochs of data were obtained before the monitoring campaign was terminated due to the \emph{Swift} boost mission\footnote{https://science.nasa.gov/mission/swift/swift-boost-mission/}. Unfortunately, TDE 2025amkg was not observed by \emph{Swift}. UV data of TDE 2025alrm from the Ultraviolet/Optical Telescope (UVOT; \citealt{Roming2005}) were retrieved and processed using the standard \texttt{HEASoft} pipeline. Individual exposures were co-added with the \texttt{uvotimsum} task, and photometry was performed using \texttt{uvotsource}, adopting a circular source region with a radius of $8^{\prime\prime}$ and a concentric annular background region with inner and outer radii of $16^{\prime\prime}$ and $34^{\prime\prime}$, respectively. As the \emph{Swift}/UVOT observations began after the outburst of TDE~2025alrm, the host-galaxy reference magnitudes were estimated by fitting its spectral energy distribution (SED) and subtracted from the measured magnitudes to obtain host-subtracted UVOT light curves (see Section\ref{SED}).

\subsection{Swift/XRT}

TDE 2025alrm was observed with the X-Ray Telescope (XRT) onboard \emph{Swift}. The data were processed using the \texttt{HEASoft} tasks \texttt{XRTPIPELINE} and \texttt{XRTPRODUCTS} to generate X-ray light curves and spectra. Source counts were extracted from a circular region with a radius of $50^{\prime\prime}$ centered on the target, while the background was from an annular region with inner and outer radii of $100^{\prime\prime}$ and $250^{\prime\prime}$, respectively. No significant X-ray emission was detected in either the individual or stacked images. We therefore computed $3\sigma$ upper limits on the X-ray flux using the WebPIMMS tool\footnote{https://heasarc.gsfc.nasa.gov/cgi-bin/Tools/w3pimms/w3pimms.pl}, assuming a power-law spectrum with a photon index of $\Gamma = 1.75$ \citep{Ricci2017}. From the stacked images, we derive an unabsorbed $3\sigma$ upper limit in the 0.3–10~keV band of $4.34 \times 10^{-13}~\mathrm{erg\ s^{-1}\ cm^{-2}}$, corresponding to a luminosity of $3.90 \times 10^{43}~\mathrm{erg\ s^{-1}}$.

\subsection{Einstein Probe/FXT}

We triggered X-ray follow-up observations with the Follow-up X-ray Telescope (FXT) onboard the Einstein Probe (EP; \citealt{Yuan2025EP}). One observation of TDE 2025alrm was obtained on 2026 April 10 with an exposure time of 2498 s. For TDE 2025amkg, two observations were obtained on 2026 April 8 and April 10, with exposure times of 1298 s and 2029 s, respectively. We processed the data using the Follow-up X-ray Telescope Data Analysis Software (FXTDAS), employing the FXTCHAIN task to generate light curves and spectra. Source counts were extracted from circular apertures with radii of $50^{\prime\prime}$ and $60^{\prime\prime}$ centered on TDE 2025alrm and TDE 2025amkg, respectively, while the background was taken from nearby source-free circular regions with a radius of $180^{\prime\prime}$. Similarly to the XRT observations, no significant X-ray emission was detected from either source. Using the FXT Observation Simulator\footnote{https://epfxt.ihep.ac.cn/simulation}, assuming a power-law spectrum with a photon index of $\Gamma = 1.75$ \citep{Ricci2017}, we derive unabsorbed $3\sigma$ upper limits on the X-ray fluxes of $7.34 \times 10^{-14}$ and $9.00 \times 10^{-14}~\mathrm{erg\ s^{-1}\ cm^{-2}}$, corresponding to luminosities of $6.59 \times 10^{42}$ and $8.08 \times 10^{42}~\mathrm{erg\ s^{-1}}$, respectively.

\section{Analysis and Results}
\label{analysis}

\begin{figure}
\centering
\begin{minipage}{0.5\textwidth}
\centering{\includegraphics[width=1\textwidth]{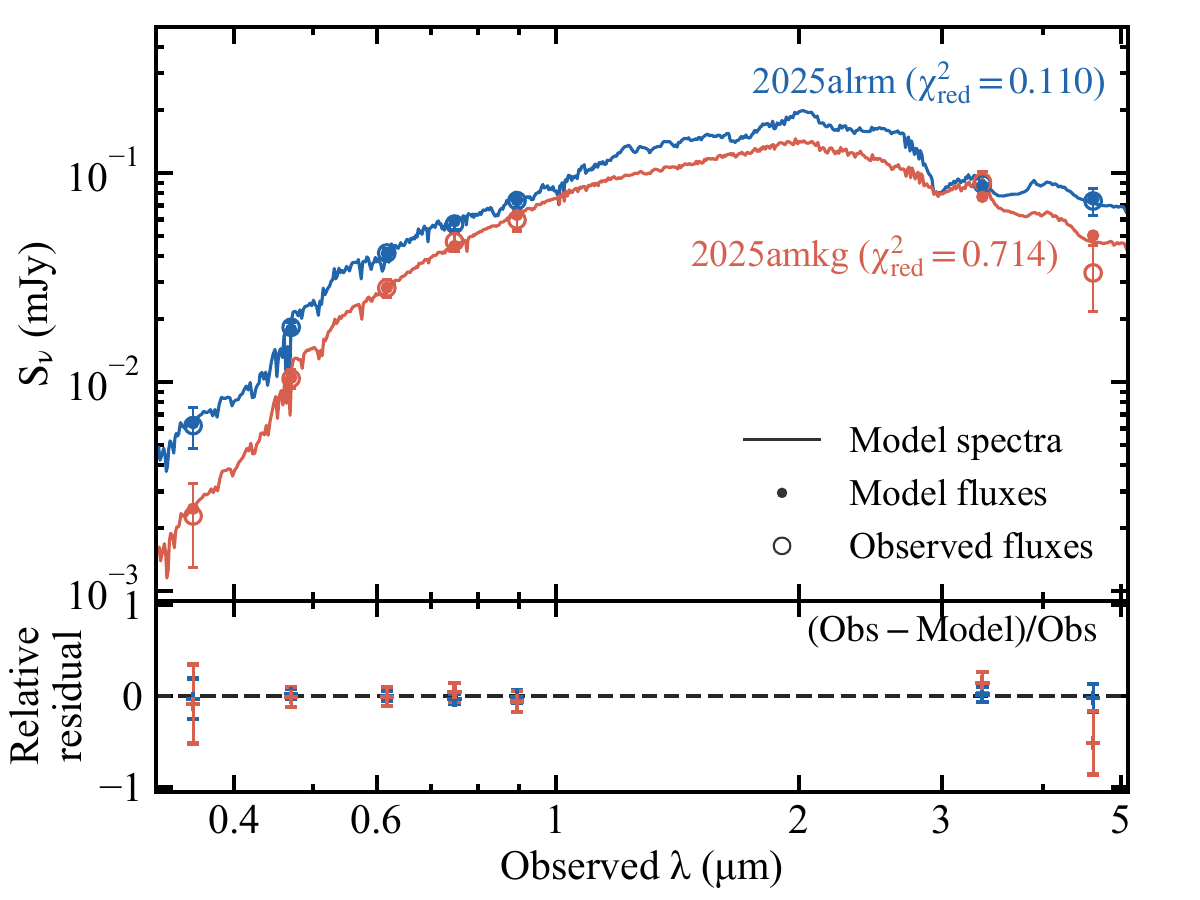}}
\end{minipage}
\caption{The host SED fitting results using the package CIGALE. Top panel: Best-fit model spectra for the host galaxies. The filled points represent the model flux for each band derived from the best-fit SED, and the open circles indicate the observed data. Bottom panel: The residuals between the observed data and model flux.}
\label{cigale}
\end{figure}

\subsection{SED Fitting of the Host Galaxy}
\label{SED}
Because no spectroscopic observations of the two sources were obtained prior to the flare and the flares are still ongoing, we are unable to acquire a pure host-galaxy spectrum. Therefore, we construct the SED using archival photometry from the Sloan Digital Sky Survey DR6 (SDSS; \citealt{SDSSDR6}) and the AllWISE catalog \citep{allwise}. After correcting for the Galactic extinction, we perform SED fitting with the Code Investigating Galaxy Emission (CIGALE; \citealt{cigale}). We adopt a delayed star formation history with an optional exponential burst, the single stellar population of \citet{Bruzual2003}, a modified \citet{Calzetti2000} dust attenuation law, and a \citet{dale2014} dust emission component. In our fitting (see Figure \ref{cigale}), the derived stellar masses of the host galaxies of TDE 2025alrm and TDE 2025amkg are $(6.89 \pm 0.61) \times 10^{9}$~\msun\ and $(7.35 \pm 1.35) \times 10^{9}$~\msun, respectively. Based on the relation between the black hole and stellar mass presented in \citet{Reines2015}, we derive black hole masses of $10^{6.23}$~\msun and $10^{6.26}$~\msun, which are consistent with the black hole masses of typical TDE hosts.

\begin{figure}
\centering
\begin{minipage}{0.5\textwidth}
\centering{\includegraphics[width=1\textwidth]{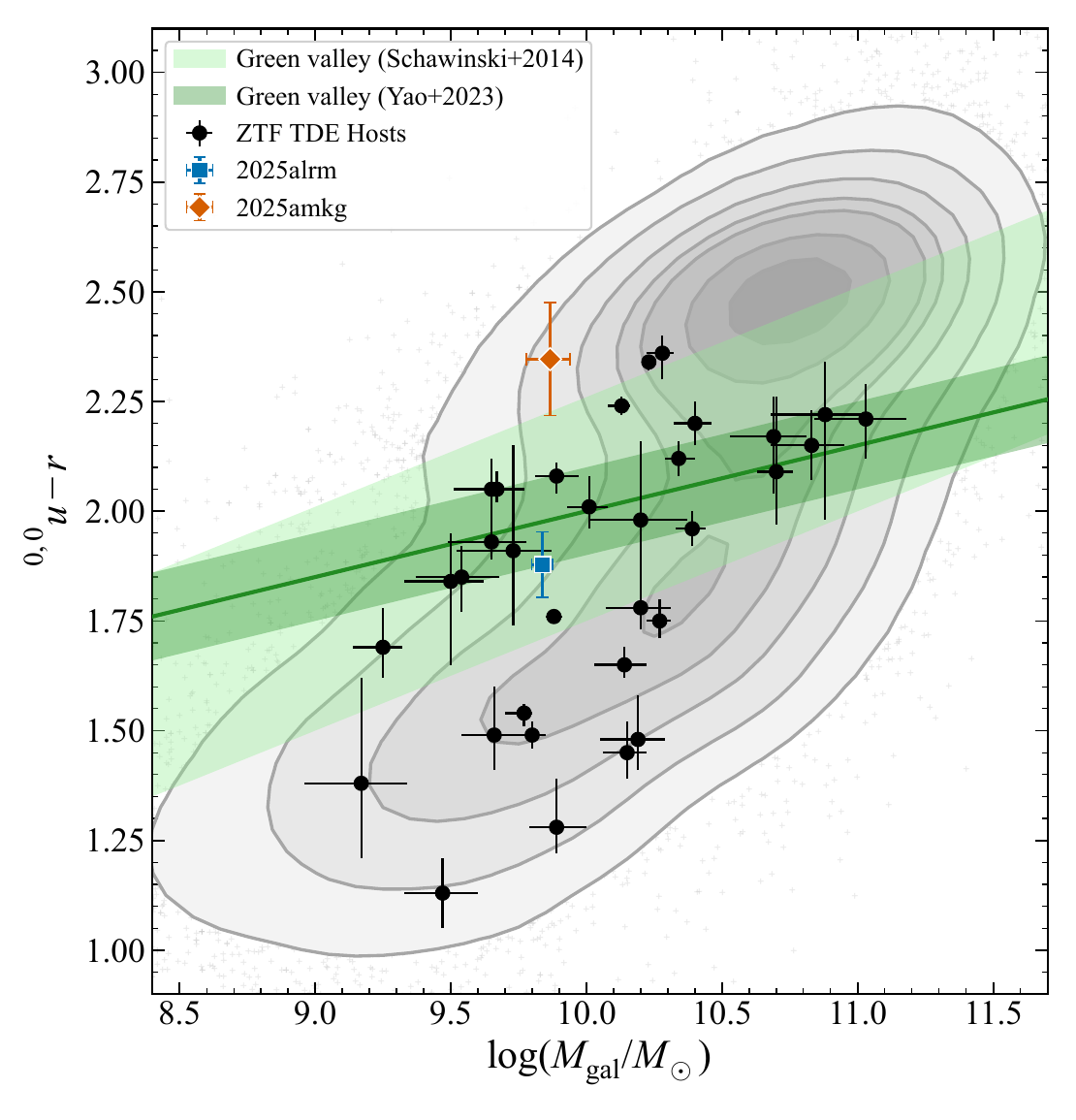}}
\end{minipage}
\caption{Host galaxies of the TDEs on the stellar mass versus rest-frame \(u-r\) diagram. The blue and orange dots denote the hosts of TDE 2025alrm and TDE 2025amkg, respectively, and the black points represent the ZTF TDE hosts. The green shaded regions indicate the green-valley definitions adopted from \citet{Schawinski2014} and \citet{Yao2023}. The gray contours show the distribution of the comparison galaxy sample.}
\label{green}
\end{figure}

To examine whether the host galaxies of TDE 2025alrm and TDE 2025amkg are consistent with the known preference of optical TDEs for green-valley galaxies \citep{Hammerstein2021}, we placed them on the stellar mass versus rest-frame \(u-r\) diagram following \citet{Yao2023}. As shown in Figure \ref{green}, the host of TDE 2025alrm lies within the green-valley region, while the host of TDE 2025amkg lies slightly above it. Nevertheless, both hosts fall within the overall distribution of known TDE hosts, indicating that their host properties are broadly consistent with those of previously identified optical TDEs.

\subsection{Photometric Analysis}
\label{photometric}

TDE~2025alrm was first detected by WFST in the $g$ and $u$ bands at MJD $\sim 60999$. The optical flux subsequently rose steadily, reached a peak at MJD $\sim 61023.8$, and then entered a decline phase. Similarly, TDE~2025amkg was first detected in the $g$ band at MJD $\sim 60970$, with the flux rising to a peak at MJD $\sim 60995.8$ before beginning to fade. To characterize the temporal behavior of the two sources, we fitted the $g$ band light curves using a rise–decline model, applying the Markov Chain Monte Carlo (MCMC) sampler \texttt{emcee} \citep{emcee}. The function is given by:

\begin{equation}
f_g(t)=f_{g,~\text{peak}}\times
\begin{cases}
    e^{[-(t-t_\text{peak})^2/(2\sigma^2)]} &t\leq t_\text{peak}\\
    (\frac{t-t_\text{peak}+t_0}{t_0})^{-5/3} &t>t_\text{peak}
\end{cases}
\label{lightcurve}
\end{equation}
We used 32 walkers, each evolved for \(10^5\) steps, and discarded the first \(9\times10^4\) steps as burn-in. Convergence was assessed using the integrated autocorrelation time \(\tau\), which was \(<33\) steps for all fitted parameters. The post-burn-in chain length therefore exceeded \(300\tau\), indicating satisfactory convergence. The best-fitting parameters for TDE~2025alrm are $\sigma=13.28^{+0.26}_{-0.26}~\text{days}$ and $t_0=72.23^{+2.46}_{-2.38}~\text{days}$. For TDE~2025amkg, the corresponding values are $\sigma=34.81^{+0.98}_{-0.91}~\text{days}$ and $t_0=220.74^{+6.53}_{-6.23}~\text{days}$. From these parameters, we derive the rest-frame rise timescale from half-maximum to peak luminosity ($t_{1/2,\text{rise}}$) and the decline timescale from peak to half-maximum ($t_{1/2,\text{decline}}$). For TDE~2025alrm, we obtain $t_{1/2,\text{rise}}=13.2~\text{days}$ and $t_{1/2,\text{decline}}=31.6~\text{days}$, while for TDE~2025amkg we obtain $t_{1/2,\text{rise}}=34.7~\text{days}$ and $t_{1/2,\text{decline}}=96.4~\text{days}$. These values fall within the typical timescale range of optically selected TDEs \citep{vV2021, Hammerstein2023, Yao2023}.

\begin{figure}
\centering
\begin{minipage}{0.5\textwidth}
\centering{\includegraphics[width=0.9\textwidth]{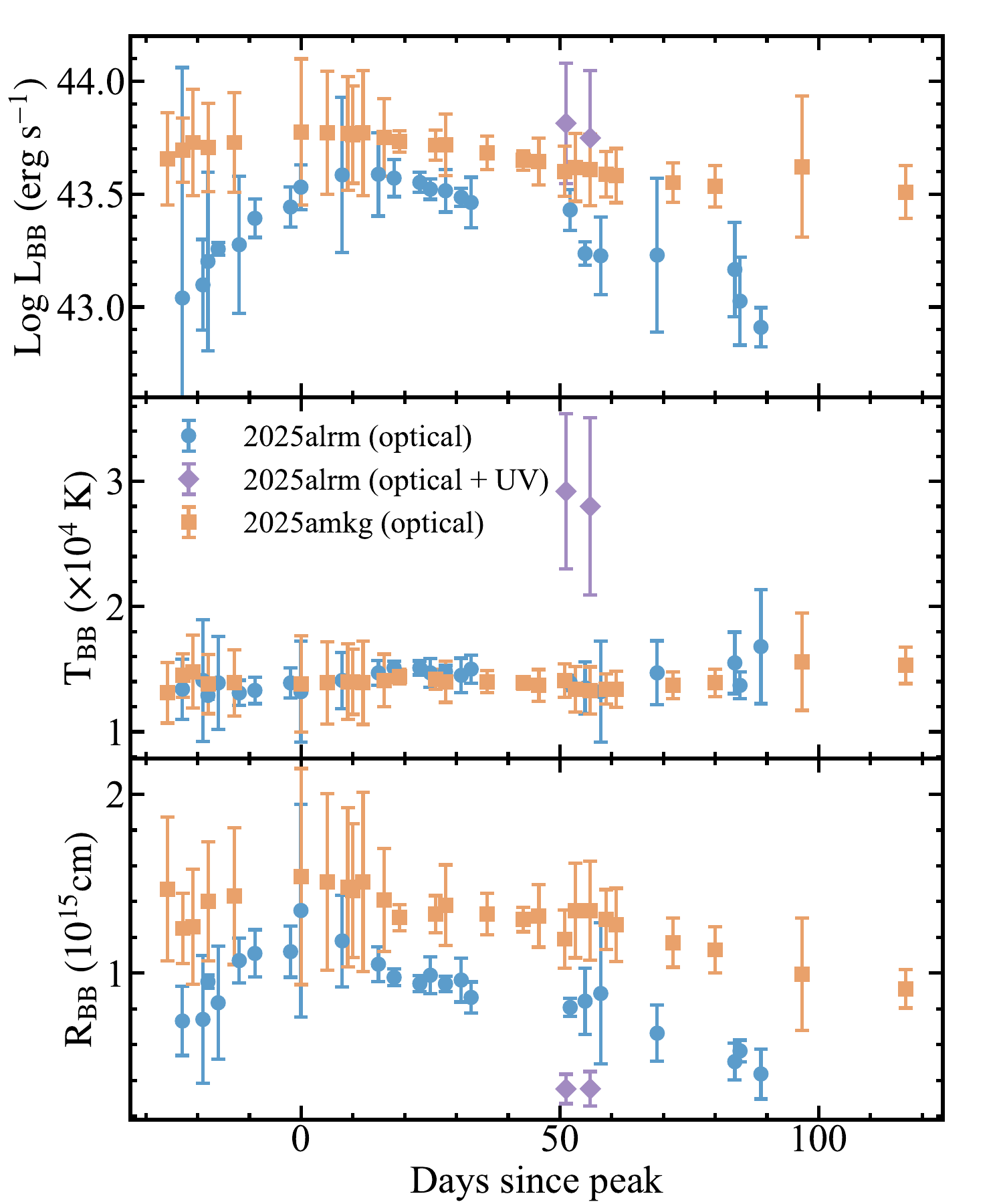}}
\end{minipage}
\caption{The fitted blackbody luminosity, blackbody temperature, and blackbody radius of TDE~2025alrm and TDE~2025amkg. Different colors denote different datasets used in the fitting. The days since peak are relative to the dates of $g$ band peaks (MJD $\sim61023.8$ and $\sim60995.8$, respectively).}
\label{superbol}
\end{figure}

In addition, we fitted the optical light curves with a blackbody SED using \texttt{Superbol} \citep{Nicholl2018}. The fitting was performed using three well sampled WFST bands ($u$, $g$ and $r$), with the $g$ band adopted as the reference band to define the fitting epochs. The $u$ and $r$ band light curves were linearly interpolated to the corresponding $g$ band epochs. At each epoch, the multi-band photometry was used to construct an SED, which was then independently fitted with a blackbody to determine the temperature ($T_{\rm BB}$), photospheric radius ($R_{\rm BB}$) and blackbody luminosity ($L_{\rm BB}$).
Both TDE~2025alrm and TDE~2025amkg exhibit approximately constant blackbody temperatures of \(\sim 14{,}000~\mathrm{K}\), with peak blackbody luminosities of $10^{43.6\pm0.2}$~\lum and $10^{43.8\pm0.3}$~\lum, respectively (see Figure \ref{superbol}). These results are consistent with the properties of typical optically selected TDEs \citep{Vv2020SSRv}. We also repeated the fit for TDE 2025alrm including both optical and UV bands, which yields a higher blackbody temperature ($\sim 30{,}000~\text{K}$) and a larger blackbody luminosity. Such discrepancies between only optical and optical+UV fits have also been reported in previous studies (e.g., \citet{Lin2025}).

\subsection{Spectroscopic Analysis}
\label{Spectroscopic}

For TDE 2025alrm, we determine a redshift of z=0.179 from Balmer absorption lines, which trace the stellar population of the host galaxy. For TDE 2025amkg, we derive a redshift of z=0.180 from Ca H$\&$K absorption lines (see Figure \ref{MMT}).
To better characterize the emission-line features and mitigate host-galaxy contamination, we decomposed the observed spectra as follows. After correcting for Galactic extinction using the dust map of \citet{Schlafly2011} and the extinction curve of \citet{Fitzpatrick1999}, we modeled the pseudo-continuum of each spectrum using the TWFIT code\footnote{https://github.com/wybustc/twfit/} \citep{Wang202418ap} as a non-negative linear combination of stellar templates and a blackbody component representing the transient emission.  
For the stellar templates, we adopted six non-negative independent components compressed from the BC03 simple stellar population library \citep{Bruzual2003} by \citet{Lu2006}. The blackbody temperatures were fixed at 30000 K for 2025alrm and 14000 K for 2025amkg, corresponding to the temperatures inferred from the multiband SED fitting. During the fitting, prominent emission-line regions and the poorly flux-calibrated region at wavelengths \(\lesssim4600\) \AA\ in the observed frame were masked.

The resulting decomposition is shown in Figure~\ref{MMT}. After subtracting the fitted stellar and blackbody continua, the residual spectra reveal broad H\(\alpha\), a blended He II \(\lambda4686\)+N III \(\lambda4640\) feature, and N III \(\lambda4100\) emission in both objects, although the He II+N III blend is relatively weak in 2025alrm. The contribution of H\(\delta\) near N III \(\lambda4100\) is expected to be small, given the weakness of H\(\gamma\) and H\(\beta\). Both objects can therefore be classified as members of the H+He spectral class of TDEs, which is also the most common spectral class in the current sample \citep{Hammerstein2023}.
We further fitted the H\(\alpha\) profiles with single-Gaussian models and obtained FWHMs of approximately 20,000 and 1,650 km s\(^{-1}\) for 2025alrm and 2025amkg, respectively. The H\(\alpha\) width of 2025alrm is consistent with those typically observed in optical TDEs, which are of order \(\sim10^4\) km s\(^{-1}\) \citep{Charalampopoulos2022}. In contrast, the H\(\alpha\) line of 2025amkg is considerably narrower. However, this spectrum was obtained at a relatively late phase, and the Balmer emission lines of TDEs are known to become narrower with time \citep{Charalampopoulos2022}.


\section{Discussion}

\subsection{The Role of $u$ Band in Early TDE Identification}
\label{uband}


The wide-field, high cadence ZTF survey in the $g$ and $r$ bands is of great success in discovering TDEs, resulting in a significantly enlarged sample over the past several years. Meanwhile, the survey has defined empirical criteria for distinguishing TDEs from contaminants, mostly supernovae (SNe), based on their $g$ and $r$ band light curves. Generally, TDEs exhibit longer rise timescales than SNe \citep{Vv2019}, and most SNe show obvious reddening after peak. It is worth noting that these criteria rely on optical light curves extending beyond peak, making it challenging to distinguish TDEs from SNe at early phases. Nevertheless, the availability of $u$-band photometry enables photometric classification of TDEs even before peak. Most TDEs maintain a nearly constant blackbody temperature of a few \(\times10^4\) K, whereas SNe typically cool rapidly after peak, resulting in TDEs maintaining bluer colors for longer timescales \citep{Vv2020SSRv,Holoien2019}. As a result, the $u-g$ and $u-r$ colors provide powerful diagnostics for separating TDEs from SNe at early times. Wu, W.Y. et al. (2026, in preparation) performed a systematic comparison of the colors of TDEs and SNe, demonstrating the effectiveness of this approach. Their results show that the early-time $g-r$ colors of TDEs and SNe are broadly similar and thus offer limited discriminatory power. In contrast, TDEs exhibit significantly bluer $u-g$ and $u-r$ colors than SNe, even at early phases, making these color indices particularly effective for early classification. As examples, the $u-g$ and $u-r$ colors of TDE 2025alrm and TDE 2025amkg remain consistently blue, with values of $\sim0$. As more TDEs and SNe are observed directly in the WFST $u$ band, in the near future the resulting dataset will provide a stronger empirical foundation for refining these color distributions and enhancing the separation of TDEs from SNe in the WFST system.

\subsection{Validation of our TDE Selection Criteria}
\label{selection}

\begin{table*}[t]
\centering
\caption{Results of the TDE Photometric Selection for WFST TDEs} 
\vspace{-5mm}
\label{table}
\begin{center}
\setlength{\tabcolsep}{6mm}{
\begin{tabular}{cccc}
\toprule
TDE name & Epochs & Reference & Selected?\\
(1)&(2)&(3)&(4)\\
\midrule
2025wet & 11 & \citet{Zhu2026} & No\\
2025cyj & >30 & \citet{cyj} & No\\
2025aapf & >30 & \citet{aapf} & Yes\\
2025afsj & 20 & \citet{afsj_concepcion,2025afsj} & No\\
2025afvr & 10 & \citet{afvr1, afvr2, afvr3} & Yes\\
2025aljd & 16 & \citet{aljd} & Yes\\
2025alrm & >20 & this work & Yes\\
2025amkg & >30 & this work & Yes\\
2026dmt & 5 & \citet{afvr3} & Yes\\
2026dta & 8 & \citet{dta} & Yes\\
2026irt & 6 & \citet{irt} & Yes\\
\bottomrule
\end{tabular}
}
\end{center}
\vspace{-1.5mm}
\begin{minipage}{1\textwidth}
\footnotesize
\textbf{Note.} Column (1): Name of TDEs well sampled by WFST. Column (2): The number of observational epochs for each TDE. Column (3): Reference papers or astronotes. Column(4): Whether the TDE satisfies our selection criteria.
\end{minipage}
\end{table*}

11 spectroscopically classified TDEs lie within the WFST surveyed area and have been well sampled from MJD 61000 (2025 November 21) to 61185 (2026 May 25). Applying the criteria described above, we successfully recovered 8 of these TDEs from the WFST database (see Table \ref{table}). The numbers of observational epochs for 2026dmt, 2026dta, and 2026irt are fewer than 10, indicating that our criteria are capable of identifying TDEs even with a limited number of observations. For 2025wet, although 11 observational epochs are available after MJD 61000, only four epochs include same-night detections in two bands, allowing color measurements. At these late phases, the source was faint, and the large photometric uncertainties caused it to fail our color selection criteria. However, when the time window is extended to cover the full light curve of 2025wet, the source can be successfully recovered. Although 2025cyj has sufficient observational coverage, its WFST $u-g$ colors are significantly redder than expected because the $u$ band reference images were obtained while the source was still in its flaring phase. If the offset in the $u$-band difference magnitudes is corrected, TDE 2025cyj would also satisfy our selection criteria. 2025afsj was not selected because it occurred in an AGN host \citep{2025afsj} and has an infrared color of $\text{W1}-\text{W2} > 0.5$. These results demonstrate that our selection criteria can effectively identify TDEs in the WFST database.

\begin{figure*}
\centering
\begin{minipage}{0.9\textwidth}
\centering{\includegraphics[width=0.9\textwidth]{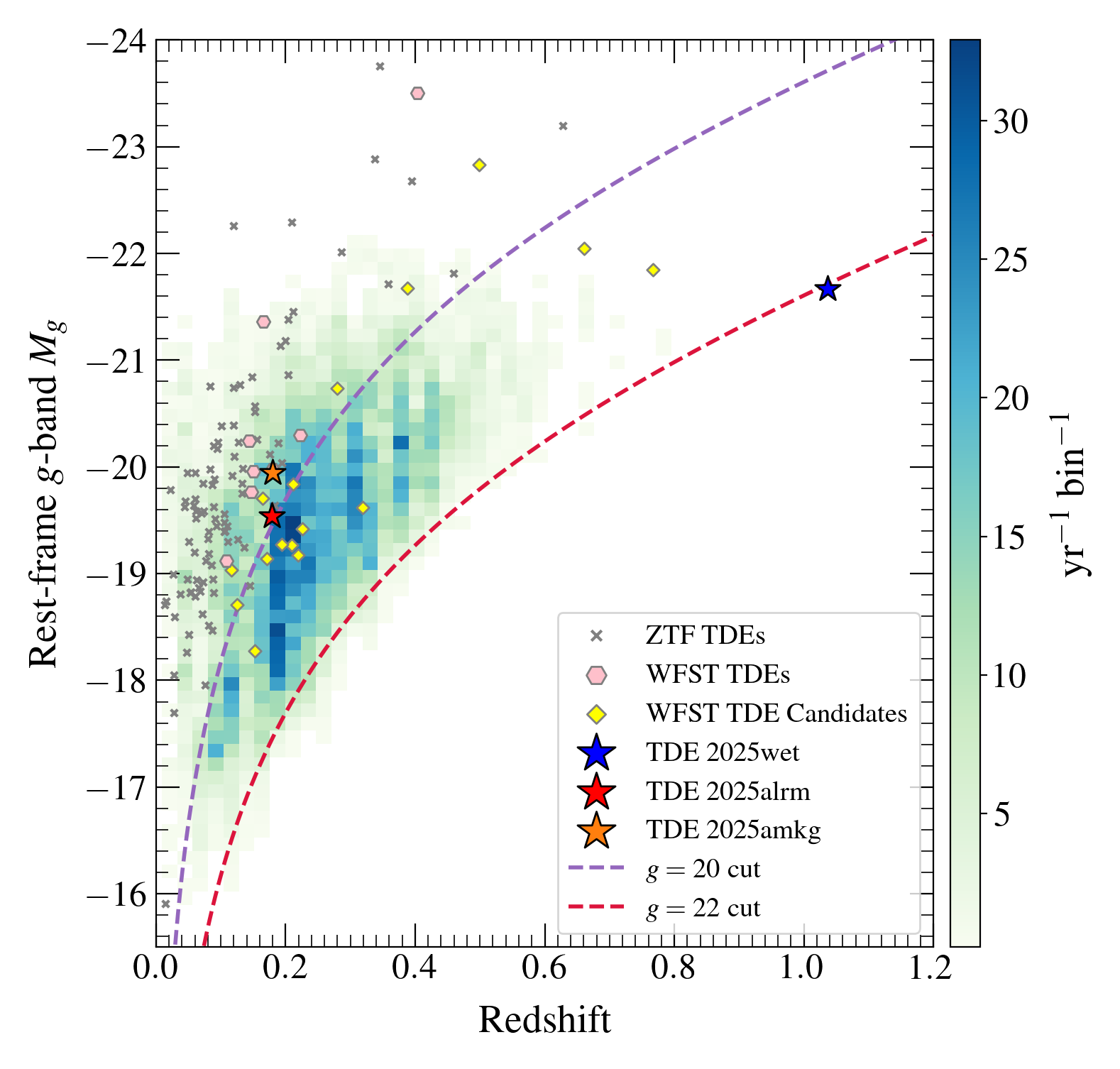}}
\end{minipage}
\caption{The distribution of redshift and peak absolute $g$ band magnitude for TDEs. The gray dots represent TDEs from the TNS, detected by ZTF during flare phases, while the pink and yellow dots denote TDEs and TDE candidates identified by WFST, respectively. The star symbols mark TDE 2025alrm, TDE 2025amkg, and TDE 2025wet. The background color map shows the simulated distribution of WFST-detectable TDEs predicted by \citet{Lin2022a}, with the color indicating the expected number of detectable TDEs per bin per year. The dashed lines in different colors denote different apparent magnitude limits.}
\label{Mgz}
\end{figure*}

Applying our selection criteria to the WFST database from MJD 61000 to 61185 yields approximately 600 sources. After excluding objects spectroscopically classified as AGNs or stars, about 300 sources remain. These sources were subsequently examined by visual inspection, resulting in a final sample of 16 TDE candidates (see Figure \ref{Mgz}). Unfortunately, due to the faintness of these sources ($g_\text{peak}\gtrsim19.5$) and the limited availability of spectroscopic resources, we were unable to obtain timely spectroscopic classifications. It is important to note that the color cuts involving the $u$ band are somewhat empirical, since systematic $u$ band observations of TDEs remain limited. Future WFST observations will build up a larger statistical sample of TDE colors in the $u$ band, providing an empirical reference for WFST and the Vera C. Rubin Observatory Legacy Survey of Space and Time (LSST; \citealt{Ivezi2019}), which will also include $u$ band observations. 

\subsection{Discovery Rate of WFST TDEs}
\label{rate}

Based on our preliminary TDE sample, we provide an initial estimate of the actual TDE discovery rate of the WFST WFS program. For this purpose, we adopt the six-month period from MJD 61000 (2025 November 21) to MJD 61185 (2026 May 25). This interval is chosen because, prior to 2025 October, WFST was primarily engaged in acquiring reference images for the WFS, and only thereafter fully transitioned into routine time-domain survey operations.
During this six-month period, WFST identified 11 spectroscopically confirmed TDEs, along with 16 additional high-confidence TDE candidates, which unfortunately lacked timely spectroscopic confirmation. Therefore, the lower limit on the TDE discovery rate is estimated to be $22\ \text{year}^{-1}$. If half of the TDE candidates are eventually confirmed as genuine TDEs, the discovery rate would increase to approximately $38\ \text{year}^{-1}$. 

The inferred TDE discovery rate is substantially lower than the prediction of \citet{Lin2022a}, who estimated a TDE discovery rate of approximately $392\pm74~ \text{year}^{-1}$. It is worth noting, however, that TDE~2025alrm, TDE~2025amkg, and most of the WFST TDE candidates occupy the highest-probability region of the predicted absolute magnitude--redshift parameter space (Figure~\ref{Mgz}), indicating that the discrepancy is unlikely to arise from the TDE luminosity function itself. Instead, the prediction of  \citet{Lin2022a} was made prior to the commissioning of WFST and therefore relied on assumptions about the survey strategy, observing conditions, and operational efficiency that proved to be overly optimistic. In practice, both the achieved survey depth and the fraction of usable observing nights have been somewhat lower than anticipated. The combined effect of these factors has naturally resulted in a TDE yield well below the original prediction. Meanwhile, the TDE selection pipeline remains under active development, and the availability of spectroscopic follow-up resources is steadily improving. Furthermore, combining our selection criteria with the Transformer-based TDE Classifier, a machine-learning approach developed for identifying TDEs in WFST data \citep{Zheng2026}, may further improve the completeness and efficiency of TDE candidate selection. With a larger TDE sample accumulated over a longer operational baseline, we expect to derive a more reliable and accurate measurement of the WFST TDE discovery rate.

\subsection{Prospects for Discovering Faint TDEs with WFST}

With observed peak magnitudes of $g\sim20$ and $g\sim19.6$, TDE 2025alrm and TDE 2025amkg are among the faintest TDEs discovered to date in terms of apparent magnitude. To further illustrate their faintness, we compare them with TDEs from the Transient Name Server (TNS) \footnote{https://www.wis-tns.org/} that were detected by ZTF during their flaring phases (see Figure~\ref{Mgz}). We applied a K-correction assuming a blackbody model with a uniform temperature of 14,000 K to derive the rest-frame $g$ band absolute magnitudes. In Figure~\ref{Mgz}, the lower-right region corresponds to fainter apparent $g$ band magnitudes. Most TDEs are brighter than the $g\sim20$ limit, whereas TDE 2025alrm and TDE 2025amkg lie close to it, and TDE 2025wet lies near the $g\sim22$ limit, highlighting the faint nature of the WFST TDEs. Only one ZTF TDE, TDE 2025lxx, is found to be fainter than the $g\sim20$ limit. However, its S/N in the $g$ band ($\sim4$) is significantly lower than those of TDE 2025alrm and TDE 2025amkg ($>20$). Furthermore, the apparent magnitudes of the WFST TDE candidates range from $g\sim19.5$ to $\sim21.5$, fainter than most TDEs. These results demonstrate the superior sensitivity of WFST and its capability for detecting faint transients. 

Detecting more faint TDEs would benefit TDE research in several important respects. First, intrinsically faint TDEs in the nearby Universe can provide valuable constraints on the faint end of the TDE luminosity function~\citep{Lin2022b,Yao2023,Zhu2023}, which is important for determining the total volumetric TDE rate. In addition, extending TDE searches to lower luminosities may increase our sensitivity to TDEs around lower-mass black holes, potentially providing new probes of the intermediate-mass black hole population (e.g.,  EP240222a, \citealt{Jin2025arXiv}). Indeed, the simulations of \citet{Lin2022a} predict that a substantial fraction ($\sim10\%$) of the TDEs discovered by WFST should have peak magnitudes as faint as $M_g>-18$ (Figure~\ref{Mgz}), a population that remains scarcely explored observationally. Unfortunately, such events have not yet been identified by WFST, most likely because of the still limited sample size ($\sim30$ TDEs). Moreover, theoretical predictions for the faint end of the TDE population remain highly uncertain, owing to our incomplete understanding of several key factors, including the SMBH occupation fraction in dwarf galaxies and the properties of TDE light curves around low-mass SMBHs. We therefore do not regard the current nondetection of this population as a significant discrepancy. Instead, it should be revisited with a substantially larger TDE sample in the coming years. 

Second, luminous TDEs at high redshift (e.g., TDE 2025wet at $z=1.037$, \citealt{Zhu2026}) will also be identified as apparently faint TDEs and of particular interest. They offer a unique opportunity since they provide direct access to the rest-frame far-ultraviolet emission, which is otherwise inaccessible and is believed to account for the long-standing "missing energy" problem in TDEs \citep{Lu2018,Zhu2026}. The prospects in this direction are especially promising, largely owing to the recent discovery of a rapidly growing new population of TDEs, namely featureless TDEs~\citep{Hammerstein2023,Yao2023,Yao2026}. These events are typically more luminous and are associated with more massive SMBHs, making them detectable out to higher redshifts. Meanwhile, discovering larger numbers of rare, luminous TDEs will place much more stringent constraints on the high end of the TDE luminosity function and on the SMBH mass function. This, in turn, will facilitate our understanding of TDEs around the most massive SMBHs, where general relativistic effects and the disruption of more massive stellar progenitors may increase the effective Hills mass and produce observable luminous flares~\citep{Gafton2019,Ryu2023,Huang2024,Mummery2024}.

Last but not least, fainter optical TDEs can be used to investigate their possible association with TDE candidates identified in other wavebands, thereby providing insights into the selection effects and limitations of TDE searches across different bands. For example, infrared-selected TDE candidates discovered in past years suggest the existence of a population of TDEs that are optically weak, likely due to dust obscuration~\citep{Jiang2021,Wang2022,Masterson2024}. The deep, wide-field surveys conducted by WFST and LSST will be essential for independently uncovering and confirming this hidden population, enabling a more complete census of TDEs and helping to mitigate the host-galaxy bias of previous optically selected TDEs~\citep{French2016}.

\section{Conclusion}

In this work, we report the pilot discovery of TDE~2025alrm and TDE~2025amkg, which have relatively faint peak apparent magnitudes of $g \sim 20.0$ and $g \sim 19.6$, respectively. Their characteristic evolution timescales, nearly constant blackbody temperatures of $\sim10^4\ \text{K}$, and broad emission features from H\(\alpha\), a blended He II \(\lambda4686\)+N III \(\lambda4640\), and N III \(\lambda4100\), together demonstrate that TDE~2025alrm and TDE~2025amkg are H+He TDEs. The two hosts are broadly consistent with the known optical TDE host population, with 2025alrm lying in the green valley and 2025amkg slightly above it. As examples, the optical light curves of these sources highlight the important role of the WFST $u$-band survey in identifying TDEs at early phases, since the $u-g$ and $u-r$ colors of TDEs are generally bluer than those of SNe. We established WFST-specific photometric selection criteria based on light-curve evolution and multi-band colors, and successfully recovered 8 of the 11 spectroscopically confirmed TDEs and discovered 16 candidates. The current WFST TDE sample implies a lower limit on the TDE discovery rate, $\sim22\ \text{year}^{-1}$, and a more optimistic rate, $\sim38\ \text{year}^{-1}$ if half of the candidates are genuine TDEs. The current WFST TDE and candidate sample demonstrates the superior sensitivity of WFST and confirms its capability to discover more apparently faint TDEs in the future, particularly low-luminosity, high-redshift, and dust-obscured events, which remain rare in current optical samples and will help expand the observed parameter space of the TDE population. The anticipated TDE sample from WFST will thus provide a much more complete census of dormant SMBHs and help address key open questions in the field, such as the origin of the optical emission and the unexpected preference for post-starburst galaxies.

\section{ACKNOWLEDGMENTS}
We thank the anonymous referee for the valuable comments, which have significantly improved the manuscript. This work is supported by the National Key Research and Development Program of China (2023YFA1608100), the Strategic Priority Research Program of the Chinese Academy of Sciences (XDB0550200), the National Natural Science Foundation of China (grants 12522303,12192221,12393814). The authors appreciate the support of the Cyrus Chun Ying Tang Foundations.  The Wide Field Survey Telescope (WFST) is a joint facility of the University of Science and Technology of China and Purple Mountain Observatory.




\bibliographystyle{aasjournal}
\bibliography{reference}

@ARTICLE{Fabricant2019,
       author = {{Fabricant}, Daniel and {Fata}, Robert and {Epps}, Harland and {Gauron}, Thomas and {Mueller}, Mark and {Zajac}, Joseph and {Amato}, Stephen and {Barberis}, Jack and {Bergner}, Henry and {Brennan}, Patricia and {Brown}, Warren and {Chilingarian}, Igor and {Geary}, John and {Kradinov}, Vladimir and {McLeod}, Brian and {Smith}, Matthew and {Woods}, Deborah},
        title = "{Binospec: A Wide-field Imaging Spectrograph for the MMT}",
      journal = {\pasp},
         year = 2019,
        month = jul,
       volume = {131},
       number = {1001},
        pages = {075004},
          doi = {10.1088/1538-3873/ab1d78},
archivePrefix = {arXiv},
       eprint = {1905.03320},
 primaryClass = {astro-ph.IM},
       adsurl = {https://ui.adsabs.harvard.edu/abs/2019PASP..131g5004F}
}

@ARTICLE{Masterson2024,
       author = {{Masterson}, Megan and {De}, Kishalay and {Panagiotou}, Christos and {Kara}, Erin and {Arcavi}, Iair and {Eilers}, Anna-Christina and {Frostig}, Danielle and {Gezari}, Suvi and {Grotova}, Iuliia and {Liu}, Zhu and {Malyali}, Adam and {Meisner}, Aaron M. and {Merloni}, Andrea and {Newsome}, Megan and {Rau}, Arne and {Simcoe}, Robert A. and {van Velzen}, Sjoert},
        title = "{A New Population of Mid-infrared-selected Tidal Disruption Events: Implications for Tidal Disruption Event Rates and Host Galaxy Properties}",
      journal = {\apj},
         year = 2024,
        month = feb,
       volume = {961},
       number = {2},
          eid = {211},
        pages = {211},
          doi = {10.3847/1538-4357/ad18bb},
archivePrefix = {arXiv},
       eprint = {2401.01403},
 primaryClass = {astro-ph.HE},
       adsurl = {https://ui.adsabs.harvard.edu/abs/2024ApJ...961..211M}
}

@ARTICLE{Lu2018,
       author = {{Lu}, Wenbin and {Kumar}, Pawan},
        title = "{On the Missing Energy Puzzle of Tidal Disruption Events}",
      journal = {\apj},
         year = 2018,
        month = oct,
       volume = {865},
       number = {2},
          eid = {128},
        pages = {128},
          doi = {10.3847/1538-4357/aad54a},
archivePrefix = {arXiv},
       eprint = {1802.02151},
 primaryClass = {astro-ph.HE},
       adsurl = {https://ui.adsabs.harvard.edu/abs/2018ApJ...865..128L}
}

@ARTICLE{Wang2022,
       author = {{Wang}, Yibo and {Jiang}, Ning and {Wang}, Tinggui and {Zhu}, Jiazheng and {Dou}, Liming and {Lin}, Zheyu and {Sun}, Luming and {Liu}, Hui and {Sheng}, Zhenfeng},
        title = "{Discovery of ATLAS17jrp as an Optical-, X-Ray-, and Infrared-bright Tidal Disruption Event in a Star-forming Galaxy}",
      journal = {\apjl},
         year = 2022,
        month = may,
       volume = {930},
       number = {1},
          eid = {L4},
        pages = {L4},
          doi = {10.3847/2041-8213/ac6670},
archivePrefix = {arXiv},
       eprint = {2204.05461},
 primaryClass = {astro-ph.GA},
       adsurl = {https://ui.adsabs.harvard.edu/abs/2022ApJ...930L...4W}
}

@ARTICLE{Yuan2025EP,
       author = {{Yuan}, Weimin and {Dai}, Lixin and {Feng}, Hua and {Jin}, Chichuan and {Jonker}, Peter and {Kuulkers}, Erik and {Liu}, Yuan and {Nandra}, Kirpal and {O'Brien}, Paul and {Piro}, Luigi and {Rau}, Arne and {Rea}, Nanda and {Sanders}, Jeremy and {Tao}, Lian and {Wang}, Junfeng and {Wu}, Xuefeng and {Zhang}, Bing and {Zhang}, Shuangnan and {Ai}, Shunke and {Buchner}, Johannes and {Bulbul}, Esra and {Chen}, Hechao and {Chen}, Minghua and {Chen}, Yong and {Chen}, Yu-Peng and {Coleiro}, Alexis and {Coti Zelati}, Francesco and {Dai}, Zigao and {Fan}, Xilong and {Fan}, Zhou and {Friedrich}, Susanne and {Gao}, He and {Ge}, Chong and {Ge}, Mingyu and {Geng}, Jinjun and {Ghirlanda}, Giancarlo and {Gianfagna}, Giulia and {Gou}, Lijun and {Guillot}, S{\'e}bastien and {Hou}, Xian and {Hu}, Jingwei and {Huang}, Yongfeng and {Ji}, Long and {Jia}, Shumei and {Komossa}, S. and {Kong}, Albert K.~H. and {Lan}, Lin and {Li}, An and {Li}, Ang and {Li}, Chengkui and {Li}, Dongyue and {Li}, Jian and {Li}, Zhaosheng and {Ling}, Zhixing and {Liu}, Ang and {Liu}, Jinzhong and {Liu}, Liangduan and {Liu}, Zhu and {Luo}, Jiawei and {Ma}, Ruican and {Maggi}, Pierre and {Maitra}, Chandreyee and {Marino}, Alessio and {Chi-Yung Ng}, Stephen and {Pan}, Haiwu and {Rukdee}, Surangkhana and {Soria}, Roberto and {Sun}, Hui and {Tam}, Pak-Hin Thomas and {Linesh Thakur}, Aishwarya and {Tian}, Hui and {Troja}, Eleonora and {Wang}, Wei and {Wang}, Xiangyu and {Wang}, Yanan and {Wei}, Junjie and {Wen}, Sixiang and {Wu}, Jianfeng and {Wu}, Ting and {Xiao}, Di and {Xu}, Dong and {Xu}, Renxin and {Xu}, Yanjun and {Xu}, Yu and {Yang}, Haonan and {You}, Bei and {Yu}, Heng and {Yu}, Yunwei and {Zhang}, Binbin and {Zhang}, Chen and {Zhang}, Guobao and {Zhang}, Liang and {Zhang}, Wenda and {Zhang}, Yu and {Zhou}, Ping and {Zou}, Zecheng},
        title = "{Science objectives of the Einstein Probe mission}",
      journal = {arXiv e-prints},
         year = 2025,
        month = jan,
          eid = {arXiv:2501.07362},
        pages = {arXiv:2501.07362},
          doi = {10.48550/arXiv.2501.07362},
archivePrefix = {arXiv},
       eprint = {2501.07362},
 primaryClass = {astro-ph.HE},
       adsurl = {https://ui.adsabs.harvard.edu/abs/2025arXiv250107362Y}
}

@ARTICLE{vV2021,
       author = {{van Velzen}, Sjoert and {Gezari}, Suvi and {Hammerstein}, Erica and {Roth}, Nathaniel and {Frederick}, Sara and {Ward}, Charlotte and {Hung}, Tiara and {Cenko}, S. Bradley and {Stein}, Robert and {Perley}, Daniel A. and {Taggart}, Kirsty and {Foley}, Ryan J. and {Sollerman}, Jesper and {Blagorodnova}, Nadejda and {Andreoni}, Igor and {Bellm}, Eric C. and {Brinnel}, Valery and {De}, Kishalay and {Dekany}, Richard and {Feeney}, Michael and {Fremling}, Christoffer and {Giomi}, Matteo and {Golkhou}, V. Zach and {Graham}, Matthew J. and {Ho}, Anna. Y.~Q. and {Kasliwal}, Mansi M. and {Kilpatrick}, Charles D. and {Kulkarni}, Shrinivas R. and {Kupfer}, Thomas and {Laher}, Russ R. and {Mahabal}, Ashish and {Masci}, Frank J. and {Miller}, Adam A. and {Nordin}, Jakob and {Riddle}, Reed and {Rusholme}, Ben and {van Santen}, Jakob and {Sharma}, Yashvi and {Shupe}, David L. and {Soumagnac}, Maayane T.},
        title = "{Seventeen Tidal Disruption Events from the First Half of ZTF Survey Observations: Entering a New Era of Population Studies}",
      journal = {\apj},
         year = 2021,
        month = feb,
       volume = {908},
       number = {1},
          eid = {4},
        pages = {4},
          doi = {10.3847/1538-4357/abc258},
archivePrefix = {arXiv},
       eprint = {2001.01409},
 primaryClass = {astro-ph.HE},
       adsurl = {https://ui.adsabs.harvard.edu/abs/2021ApJ...908....4V}
}

@ARTICLE{French2016,
       author = {{French}, K. Decker and {Arcavi}, Iair and {Zabludoff}, Ann},
        title = "{Tidal Disruption Events Prefer Unusual Host Galaxies}",
      journal = {\apjl},
         year = 2016,
        month = feb,
       volume = {818},
       number = {1},
          eid = {L21},
        pages = {L21},
          doi = {10.3847/2041-8205/818/1/L21},
archivePrefix = {arXiv},
       eprint = {1601.04705},
 primaryClass = {astro-ph.GA},
       adsurl = {https://ui.adsabs.harvard.edu/abs/2016ApJ...818L..21F}
}

@ARTICLE{Gezari2021,
       author = {{Gezari}, Suvi},
        title = "{Tidal Disruption Events}",
      journal = {\araa},
         year = 2021,
        month = sep,
       volume = {59},
        pages = {21-58},
          doi = {10.1146/annurev-astro-111720-030029},
archivePrefix = {arXiv},
       eprint = {2104.14580},
 primaryClass = {astro-ph.HE},
       adsurl = {https://ui.adsabs.harvard.edu/abs/2021ARA&A..59...21G}
}

@ARTICLE{Yao2023,
       author = {{Yao}, Yuhan and {Ravi}, Vikram and {Gezari}, Suvi and {van Velzen}, Sjoert and {Lu}, Wenbin and {Schulze}, Steve and {Somalwar}, Jean J. and {Kulkarni}, S.~R. and {Hammerstein}, Erica and {Nicholl}, Matt and {Graham}, Matthew J. and {Perley}, Daniel A. and {Cenko}, S. Bradley and {Stein}, Robert and {Ricarte}, Angelo and {Chadayammuri}, Urmila and {Quataert}, Eliot and {Bellm}, Eric C. and {Bloom}, Joshua S. and {Dekany}, Richard and {Drake}, Andrew J. and {Groom}, Steven L. and {Mahabal}, Ashish A. and {Prince}, Thomas A. and {Riddle}, Reed and {Rusholme}, Ben and {Sharma}, Yashvi and {Sollerman}, Jesper and {Yan}, Lin},
        title = "{Tidal Disruption Event Demographics with the Zwicky Transient Facility: Volumetric Rates, Luminosity Function, and Implications for the Local Black Hole Mass Function}",
      journal = {\apjl},
         year = 2023,
        month = sep,
       volume = {955},
       number = {1},
          eid = {L6},
        pages = {L6},
          doi = {10.3847/2041-8213/acf216},
archivePrefix = {arXiv},
       eprint = {2303.06523},
 primaryClass = {astro-ph.HE},
       adsurl = {https://ui.adsabs.harvard.edu/abs/2023ApJ...955L...6Y}
}

@ARTICLE{Hammerstein2023,
       author = {{Hammerstein}, Erica and {van Velzen}, Sjoert and {Gezari}, Suvi and {Cenko}, S. Bradley and {Yao}, Yuhan and {Ward}, Charlotte and {Frederick}, Sara and {Villanueva}, Natalia and {Somalwar}, Jean J. and {Graham}, Matthew J. and {Kulkarni}, Shrinivas R. and {Stern}, Daniel and {Andreoni}, Igor and {Bellm}, Eric C. and {Dekany}, Richard and {Dhawan}, Suhail and {Drake}, Andrew J. and {Fremling}, Christoffer and {Gatkine}, Pradip and {Groom}, Steven L. and {Ho}, Anna Y.~Q. and {Kasliwal}, Mansi M. and {Karambelkar}, Viraj and {Kool}, Erik C. and {Masci}, Frank J. and {Medford}, Michael S. and {Perley}, Daniel A. and {Purdum}, Josiah and {van Roestel}, Jan and {Sharma}, Yashvi and {Sollerman}, Jesper and {Taggart}, Kirsty and {Yan}, Lin},
        title = "{The Final Season Reimagined: 30 Tidal Disruption Events from the ZTF-I Survey}",
      journal = {\apj},
         year = 2023,
        month = jan,
       volume = {942},
       number = {1},
          eid = {9},
        pages = {9},
          doi = {10.3847/1538-4357/aca283},
archivePrefix = {arXiv},
       eprint = {2203.01461},
 primaryClass = {astro-ph.HE},
       adsurl = {https://ui.adsabs.harvard.edu/abs/2023ApJ...942....9H}
}

@ARTICLE{emcee,
       author = {{Foreman-Mackey}, Daniel and {Hogg}, David W. and {Lang}, Dustin and {Goodman}, Jonathan},
        title = "{emcee: The MCMC Hammer}",
      journal = {\pasp},
         year = 2013,
        month = mar,
       volume = {125},
       number = {925},
        pages = {306},
          doi = {10.1086/670067},
archivePrefix = {arXiv},
       eprint = {1202.3665},
 primaryClass = {astro-ph.IM},
       adsurl = {https://ui.adsabs.harvard.edu/abs/2013PASP..125..306F}
}

@ARTICLE{Linjie2025,
       author = {{Lin}, Jie and {Wang}, Tinggui and {Cai}, Minxuan and {Wan}, Zhen and {Li}, Xuzhi and {Fan}, Lulu and {Zhu}, Qingfeng and {Jiang}, Ji-an and {Jiang}, Ning and {Kong}, Xu and {Lin}, Zheyu and {Zhu}, Jiazheng and {Liu}, Zhengyan and {Gao}, Jie and {Li}, Bin and {Li}, Feng and {Liang}, Ming and {Liu}, Hao and {Liu}, Wei and {Luo}, Wentao and {Tang}, Jinlong and {Wang}, Hairen and {Wang}, Jian and {Xue}, Yongquan and {Yao}, Dazhi and {Zhang}, Hongfei and {Zhang}, Xiaoling and {Zhao}, Wen and {Zheng}, Xianzhong},
        title = "{Minute-cadence Observations of the Galactic Plane with the Wide Field Survey Telescope (WFST): Overview, Methodology, and Early Results}",
      journal = {\apjs},
         year = 2025,
        month = may,
       volume = {278},
       number = {1},
          eid = {29},
        pages = {29},
          doi = {10.3847/1538-4365/adc0fc},
archivePrefix = {arXiv},
       eprint = {2412.12601},
 primaryClass = {astro-ph.SR},
       adsurl = {https://ui.adsabs.harvard.edu/abs/2025ApJS..278...29L}
}

@ARTICLE{Yao2026,
       author = {{Yao}, Yuhan and {Chornock}, Ryan and {Mummery}, Andrew and {Margutti}, Raffaella and {Gilfanov}, Marat and {Guolo}, Muryel and {Coughlin}, Eric R. and {Lu}, Wenbin and {Chakraborty}, Joheen and {Pasham}, Dheeraj R. and {Alexander}, Kate D. and {Aspegren}, Olivia and {Angus}, Charlotte R. and {Guo}, Xinze and {Hall}, Xander J. and {Hammerstein}, Erica and {Hinds}, K.-Ryan and {Ho}, Anna Y.~Q. and {Huang}, Xiaoshan and {Kammoun}, Elias and {LeBaron}, Natalie and {Lucchini}, Matteo and {McGrath}, Zo{\"e} and {Nicholl}, Matt and {Perley}, Daniel A. and {Rich}, R. Michael and {Schroeder}, Genevieve and {Sheng}, Xinyue and {Sollerman}, Jesper and {Somalwar}, Jean and {Wise}, Jacob L. and {Coughlin}, Michael W. and {Drake}, Andrew and {Graham}, Matthew J. and {Helou}, George and {Jaimes}, Joahan C. and {Kasliwal}, Mansi M. and {Mahabal}, Ashish A. and {Medvedev}, Pavel and {Purdum}, Josiah and {Rusholme}, Ben and {Sunyaev}, Rashid},
        title = "{AT2024lhc and AT2024kmq in the landscape of featureless tidal disruption events}",
      journal = {\mnras},
         year = 2026,
        month = jun,
       volume = {549},
       number = {2},
          eid = {stag920},
        pages = {stag920},
          doi = {10.1093/mnras/stag920},
archivePrefix = {arXiv},
       eprint = {2602.21624},
 primaryClass = {astro-ph.HE},
       adsurl = {https://ui.adsabs.harvard.edu/abs/2026MNRAS.549ag920Y}
}

@ARTICLE{Lin2022a,
       author = {{Lin}, Zheyu and {Jiang}, Ning and {Kong}, Xu},
        title = "{The prospects of finding tidal disruption events with 2.5-m Wide-Field Survey Telescope based on mock observations}",
      journal = {\mnras},
         year = 2022,
        month = jun,
       volume = {513},
       number = {2},
        pages = {2422-2436},
          doi = {10.1093/mnras/stac946},
archivePrefix = {arXiv},
       eprint = {2204.01615},
 primaryClass = {astro-ph.HE},
       adsurl = {https://ui.adsabs.harvard.edu/abs/2022MNRAS.513.2422L}
}

@ARTICLE{Lin2022b,
       author = {{Lin}, Zheyu and {Jiang}, Ning and {Kong}, Xu and {Huang}, Shifeng and {Lin}, Zesen and {Zhu}, Jiazheng and {Wang}, Yibo},
        title = "{The Luminosity Function of Tidal Disruption Flares for the ZTF-I Survey}",
      journal = {\apjl},
         year = 2022,
        month = nov,
       volume = {939},
       number = {2},
          eid = {L33},
        pages = {L33},
          doi = {10.3847/2041-8213/ac9c63},
archivePrefix = {arXiv},
       eprint = {2210.14950},
 primaryClass = {astro-ph.HE},
       adsurl = {https://ui.adsabs.harvard.edu/abs/2022ApJ...939L..33L}
}

@ARTICLE{Rees1988,
       author = {{Rees}, Martin J.},
        title = "{Tidal disruption of stars by black holes of {}10$^{6}$-{}10$^{8}$ solar masses in nearby galaxies}",
      journal = {\nat},
         year = 1988,
        month = jun,
       volume = {333},
       number = {6173},
        pages = {523-528},
          doi = {10.1038/333523a0},
       adsurl = {https://ui.adsabs.harvard.edu/abs/1988Natur.333..523R}
}

@ARTICLE{Roming2005,
       author = {{Roming}, Peter W.~A. and {Kennedy}, Thomas E. and {Mason}, Keith O. and {Nousek}, John A. and {Ahr}, Lindy and {Bingham}, Richard E. and {Broos}, Patrick S. and {Carter}, Mary J. and {Hancock}, Barry K. and {Huckle}, Howard E. and {Hunsberger}, S.~D. and {Kawakami}, Hajime and {Killough}, Ronnie and {Koch}, T. Scott and {McLelland}, Michael K. and {Smith}, Kelly and {Smith}, Philip J. and {Soto}, Juan Carlos and {Boyd}, Patricia T. and {Breeveld}, Alice A. and {Holland}, Stephen T. and {Ivanushkina}, Mariya and {Pryzby}, Michael S. and {Still}, Martin D. and {Stock}, Joseph},
        title = "{The Swift Ultra-Violet/Optical Telescope}",
      journal = {\ssr},
         year = 2005,
        month = oct,
       volume = {120},
       number = {3-4},
        pages = {95-142},
          doi = {10.1007/s11214-005-5095-4},
archivePrefix = {arXiv},
       eprint = {astro-ph/0507413},
 primaryClass = {astro-ph},
       adsurl = {https://ui.adsabs.harvard.edu/abs/2005SSRv..120...95R}
}

@ARTICLE{Masci2019,
       author = {{Masci}, Frank J. and {Laher}, Russ R. and {Rusholme}, Ben and {Shupe}, David L. and {Groom}, Steven and {Surace}, Jason and {Jackson}, Edward and {Monkewitz}, Serge and {Beck}, Ron and {Flynn}, David and {Terek}, Scott and {Landry}, Walter and {Hacopians}, Eugean and {Desai}, Vandana and {Howell}, Justin and {Brooke}, Tim and {Imel}, David and {Wachter}, Stefanie and {Ye}, Quan-Zhi and {Lin}, Hsing-Wen and {Cenko}, S. Bradley and {Cunningham}, Virginia and {Rebbapragada}, Umaa and {Bue}, Brian and {Miller}, Adam A. and {Mahabal}, Ashish and {Bellm}, Eric C. and {Patterson}, Maria T. and {Juri{\'c}}, Mario and {Golkhou}, V. Zach and {Ofek}, Eran O. and {Walters}, Richard and {Graham}, Matthew and {Kasliwal}, Mansi M. and {Dekany}, Richard G. and {Kupfer}, Thomas and {Burdge}, Kevin and {Cannella}, Christopher B. and {Barlow}, Tom and {Van Sistine}, Angela and {Giomi}, Matteo and {Fremling}, Christoffer and {Blagorodnova}, Nadejda and {Levitan}, David and {Riddle}, Reed and {Smith}, Roger M. and {Helou}, George and {Prince}, Thomas A. and {Kulkarni}, Shrinivas R.},
        title = "{The Zwicky Transient Facility: Data Processing, Products, and Archive}",
      journal = {\pasp},
         year = 2019,
        month = jan,
       volume = {131},
       number = {995},
        pages = {018003},
          doi = {10.1088/1538-3873/aae8ac},
archivePrefix = {arXiv},
       eprint = {1902.01872},
 primaryClass = {astro-ph.IM},
       adsurl = {https://ui.adsabs.harvard.edu/abs/2019PASP..131a8003M}
}

@ARTICLE{Wang2023,
       author = {{Wang}, Tinggui and {Liu}, Guilin and {Cai}, Zhenyi and {Geng}, Jinjun and {Fang}, Min and {He}, Haoning and {Jiang}, Ji-an and {Jiang}, Ning and {Kong}, Xu and {Li}, Bin and {Li}, Ye and {Luo}, Wentao and {Pan}, Zhizheng and {Wu}, Xuefeng and {Yang}, Ji and {Yu}, Jiming and {Zheng}, Xianzhong and {Zhu}, Qingfeng and {Cai}, Yi-Fu and {Chen}, Yuanyuan and {Chen}, Zhiwei and {Dai}, Zigao and {Fan}, Lulu and {Fan}, Yizhong and {Fang}, Wenjuan and {He}, Zhicheng and {Hu}, Lei and {Hu}, Maokai and {Jin}, Zhiping and {Jiang}, Zhibo and {Li}, Guoliang and {Li}, Fan and {Li}, Xuzhi and {Liang}, Runduo and {Lin}, Zheyu and {Liu}, Qingzhong and {Liu}, Wenhao and {Liu}, Zhengyan and {Liu}, Wei and {Liu}, Yao and {Lou}, Zheng and {Qu}, Han and {Sheng}, Zhenfeng and {Shi}, Jianchun and {Shu}, Yiping and {Su}, Zhenbo and {Sun}, Tianrui and {Wang}, Hongchi and {Wang}, Huiyuan and {Wang}, Jian and {Wang}, Junxian and {Wei}, Daming and {Wei}, Junjie and {Xue}, Yongquan and {Yan}, Jingzhi and {Yang}, Chao and {Yuan}, Ye and {Yuan}, Yefei and {Zhang}, Hongxin and {Zhang}, Miaomiao and {Zhao}, Haibin and {Zhao}, Wen},
        title = "{Science with the 2.5-meter Wide Field Survey Telescope (WFST)}",
      journal = {Science China Physics, Mechanics, and Astronomy},
         year = 2023,
        month = oct,
       volume = {66},
       number = {10},
          eid = {109512},
        pages = {109512},
          doi = {10.1007/s11433-023-2197-5},
archivePrefix = {arXiv},
       eprint = {2306.07590},
 primaryClass = {astro-ph.IM},
       adsurl = {https://ui.adsabs.harvard.edu/abs/2023SCPMA..6609512W}
}

@ARTICLE{Cai2025,
       author = {{Cai}, Minxuan and {Xu}, Zelin and {Fan}, Lulu and {Wan}, Zhen and {Liu}, Binyang and {Kong}, Xu and {Hu}, Weida and {Hu}, Lei and {Zhu}, Qing-feng and {Li}, Guoliang and {Lin}, Jie and {Fang}, Min and {Jiang}, Ji-an and {Xue}, Yongquan and {Zhen}, Xianzhong and {Wang}, Tinggui},
        title = "{The 2.5-meter Wide Field Survey Telescope Real-time Data Processing Pipeline I: From raw data to alert distribution}",
      journal = {arXiv e-prints},
         year = 2025,
        month = jan,
          eid = {arXiv:2501.15018},
        pages = {arXiv:2501.15018},
          doi = {10.48550/arXiv.2501.15018},
archivePrefix = {arXiv},
       eprint = {2501.15018},
 primaryClass = {astro-ph.IM},
       adsurl = {https://ui.adsabs.harvard.edu/abs/2025arXiv250115018C}
}

@ARTICLE{Cardelli1989,
       author = {{Cardelli}, Jason A. and {Clayton}, Geoffrey C. and {Mathis}, John S.},
        title = "{The Relationship between Infrared, Optical, and Ultraviolet Extinction}",
      journal = {\apj},
         year = 1989,
        month = oct,
       volume = {345},
        pages = {245},
          doi = {10.1086/167900},
       adsurl = {https://ui.adsabs.harvard.edu/abs/1989ApJ...345..245C}
}

@ARTICLE{Schlafly2011,
       author = {{Schlafly}, Edward F. and {Finkbeiner}, Douglas P.},
        title = "{Measuring Reddening with Sloan Digital Sky Survey Stellar Spectra and Recalibrating SFD}",
      journal = {\apj},
         year = 2011,
        month = aug,
       volume = {737},
       number = {2},
          eid = {103},
        pages = {103},
          doi = {10.1088/0004-637X/737/2/103},
archivePrefix = {arXiv},
       eprint = {1012.4804},
 primaryClass = {astro-ph.GA},
       adsurl = {https://ui.adsabs.harvard.edu/abs/2011ApJ...737..103S}
}

@ARTICLE{Masci2023,
       author = {{Masci}, Frank J. and {Laher}, Russ R. and {Rusholme}, Benjamin and {Shupe}, David and {Paladini}, Roberta and {Groom}, Steve and {Wold}, Avery and {Miller}, Adam A. and {Drake}, Andrew},
        title = "{A New Forced Photometry Service for the Zwicky Transient Facility}",
      journal = {arXiv e-prints},
         year = 2023,
        month = may,
          eid = {arXiv:2305.16279},
        pages = {arXiv:2305.16279},
          doi = {10.48550/arXiv.2305.16279},
archivePrefix = {arXiv},
       eprint = {2305.16279},
 primaryClass = {astro-ph.IM},
       adsurl = {https://ui.adsabs.harvard.edu/abs/2023arXiv230516279M}
}

@ARTICLE{Ricci2017,
       author = {{Ricci}, C. and {Trakhtenbrot}, B. and {Koss}, M.~J. and {Ueda}, Y. and {Delvecchio}, I. and {Treister}, E. and {Schawinski}, K. and {Paltani}, S. and {Oh}, K. and {Lamperti}, I. and {Berney}, S. and {Gandhi}, P. and {Ichikawa}, K. and {Bauer}, F.~E. and {Ho}, L.~C. and {Asmus}, D. and {Beckmann}, V. and {Soldi}, S. and {Balokovi{\'c}}, M. and {Gehrels}, N. and {Markwardt}, C.~B.},
        title = "{BAT AGN Spectroscopic Survey. V. X-Ray Properties of the Swift/BAT 70-month AGN Catalog}",
      journal = {\apjs},
         year = 2017,
        month = dec,
       volume = {233},
       number = {2},
          eid = {17},
        pages = {17},
          doi = {10.3847/1538-4365/aa96ad},
archivePrefix = {arXiv},
       eprint = {1709.03989},
 primaryClass = {astro-ph.HE},
       adsurl = {https://ui.adsabs.harvard.edu/abs/2017ApJS..233...17R}
}

@ARTICLE{SDSSDR6,
       author = {{Adelman-McCarthy}, Jennifer K. and {Ag{\"u}eros}, Marcel A. and {Allam}, Sahar S. and {Allende Prieto}, Carlos and {Anderson}, Kurt S.~J. and {Anderson}, Scott F. and {Annis}, James and {Bahcall}, Neta A. and {Bailer-Jones}, C.~A.~L. and {Baldry}, Ivan K. and {Barentine}, J.~C. and {Bassett}, Bruce A. and {Becker}, Andrew C. and {Beers}, Timothy C. and {Bell}, Eric F. and {Berlind}, Andreas A. and {Bernardi}, Mariangela and {Blanton}, Michael R. and {Bochanski}, John J. and {Boroski}, William N. and {Brinchmann}, Jarle and {Brinkmann}, J. and {Brunner}, Robert J. and {Budav{\'a}ri}, Tam{\'a}s and {Carliles}, Samuel and {Carr}, Michael A. and {Castander}, Francisco J. and {Cinabro}, David and {Cool}, R.~J. and {Covey}, Kevin R. and {Csabai}, Istv{\'a}n and {Cunha}, Carlos E. and {Davenport}, James R.~A. and {Dilday}, Ben and {Doi}, Mamoru and {Eisenstein}, Daniel J. and {Evans}, Michael L. and {Fan}, Xiaohui and {Finkbeiner}, Douglas P. and {Friedman}, Scott D. and {Frieman}, Joshua A. and {Fukugita}, Masataka and {G{\"a}nsicke}, Boris T. and {Gates}, Evalyn and {Gillespie}, Bruce and {Glazebrook}, Karl and {Gray}, Jim and {Grebel}, Eva K. and {Gunn}, James E. and {Gurbani}, Vijay K. and {Hall}, Patrick B. and {Harding}, Paul and {Harvanek}, Michael and {Hawley}, Suzanne L. and {Hayes}, Jeffrey and {Heckman}, Timothy M. and {Hendry}, John S. and {Hindsley}, Robert B. and {Hirata}, Christopher M. and {Hogan}, Craig J. and {Hogg}, David W. and {Hyde}, Joseph B. and {Ichikawa}, Shin-ichi and {Ivezi{\'c}}, {\v{Z}}eljko and {Jester}, Sebastian and {Johnson}, Jennifer A. and {Jorgensen}, Anders M. and {Juri{\'c}}, Mario and {Kent}, Stephen M. and {Kessler}, R. and {Kleinman}, S.~J. and {Knapp}, G.~R. and {Kron}, Richard G. and {Krzesinski}, Jurek and {Kuropatkin}, Nikolay and {Lamb}, Donald Q. and {Lampeitl}, Hubert and {Lebedeva}, Svetlana and {Lee}, Young Sun and {French Leger}, R. and {L{\'e}pine}, S{\'e}bastien and {Lima}, Marcos and {Lin}, Huan and {Long}, Daniel C. and {Loomis}, Craig P. and {Loveday}, Jon and {Lupton}, Robert H. and {Malanushenko}, Olena and {Malanushenko}, Viktor and {Mandelbaum}, Rachel and {Margon}, Bruce and {Marriner}, John P. and {Mart{\'\i}nez-Delgado}, David and {Matsubara}, Takahiko and {McGehee}, Peregrine M. and {McKay}, Timothy A. and {Meiksin}, Avery and {Morrison}, Heather L. and {Munn}, Jeffrey A. and {Nakajima}, Reiko and {Neilsen}, Jr., Eric H. and {Newberg}, Heidi Jo and {Nichol}, Robert C. and {Nicinski}, Tom and {Nieto-Santisteban}, Maria and {Nitta}, Atsuko and {Okamura}, Sadanori and {Owen}, Russell and {Oyaizu}, Hiroaki and {Padmanabhan}, Nikhil and {Pan}, Kaike and {Park}, Changbom and {Peoples}, Jr., John and {Pier}, Jeffrey R. and {Pope}, Adrian C. and {Purger}, Norbert and {Raddick}, M. Jordan and {Re Fiorentin}, Paola and {Richards}, Gordon T. and {Richmond}, Michael W. and {Riess}, Adam G. and {Rix}, Hans-Walter and {Rockosi}, Constance M. and {Sako}, Masao and {Schlegel}, David J. and {Schneider}, Donald P. and {Schreiber}, Matthias R. and {Schwope}, Axel D. and {Seljak}, Uro{\v{s}} and {Sesar}, Branimir and {Sheldon}, Erin and {Shimasaku}, Kazu and {Sivarani}, Thirupathi and {Allyn Smith}, J. and {Snedden}, Stephanie A. and {Steinmetz}, Matthias and {Strauss}, Michael A. and {SubbaRao}, Mark and {Suto}, Yasushi and {Szalay}, Alexander S. and {Szapudi}, Istv{\'a}n and {Szkody}, Paula and {Tegmark}, Max and {Thakar}, Aniruddha R. and {Tremonti}, Christy A. and {Tucker}, Douglas L. and {Uomoto}, Alan and {Vanden Berk}, Daniel E. and {Vandenberg}, Jan and {Vidrih}, S. and {Vogeley}, Michael S. and {Voges}, Wolfgang and {Vogt}, Nicole P. and {Wadadekar}, Yogesh and {Weinberg}, David H. and {West}, Andrew A. and {White}, Simon D.~M. and {Wilhite}, Brian C. and {Yanny}, Brian and {Yocum}, D.~R. and {York}, Donald G. and {Zehavi}, Idit and {Zucker}, Daniel B.},
        title = "{The Sixth Data Release of the Sloan Digital Sky Survey}",
      journal = {\apjs},
         year = 2008,
        month = apr,
       volume = {175},
       number = {2},
        pages = {297-313},
          doi = {10.1086/524984},
archivePrefix = {arXiv},
       eprint = {0707.3413},
 primaryClass = {astro-ph},
       adsurl = {https://ui.adsabs.harvard.edu/abs/2008ApJS..175..297A}
}

@MISC{allwise,
       author = {{Cutri}, R.~M. and {Wright}, E.~L. and {Conrow}, T. and {Fowler}, J.~W. and {Eisenhardt}, P.~R.~M. and {Grillmair}, C. and {Kirkpatrick}, J.~D. and {Masci}, F. and {McCallon}, H.~L. and {Wheelock}, S.~L. and {Fajardo-Acosta}, S. and {Yan}, L. and {Benford}, D. and {Harbut}, M. and {Jarrett}, T. and {Lake}, S. and {Leisawitz}, D. and {Ressler}, M.~E. and {Stanford}, S.~A. and {Tsai}, C.~W. and {Liu}, F. and {Helou}, G. and {Mainzer}, A. and {Gettings}, D. and {Gonzalez}, A. and {Hoffman}, D. and {Marsh}, K.~A. and {Padgett}, D. and {Skrutskie}, M.~F. and {Beck}, R.~P. and {Papin}, M. and {Wittman}, M.},
        title = "{Explanatory Supplement to the AllWISE Data Release Products}",
 howpublished = {Explanatory Supplement to the AllWISE Data Release Products, by R. M. Cutri et al.},
         year = 2013,
        month = nov,
        pages = {1},
       adsurl = {https://ui.adsabs.harvard.edu/abs/2013wise.rept....1C}
}

@ARTICLE{cigale,
       author = {{Boquien}, M. and {Burgarella}, D. and {Roehlly}, Y. and {Buat}, V. and {Ciesla}, L. and {Corre}, D. and {Inoue}, A.~K. and {Salas}, H.},
        title = "{CIGALE: a python Code Investigating GALaxy Emission}",
      journal = {\aap},
         year = 2019,
        month = feb,
       volume = {622},
          eid = {A103},
        pages = {A103},
          doi = {10.1051/0004-6361/201834156},
archivePrefix = {arXiv},
       eprint = {1811.03094},
 primaryClass = {astro-ph.GA},
       adsurl = {https://ui.adsabs.harvard.edu/abs/2019A&A...622A.103B}
}

@ARTICLE{Calzetti2000,
       author = {{Calzetti}, Daniela and {Armus}, Lee and {Bohlin}, Ralph C. and {Kinney}, Anne L. and {Koornneef}, Jan and {Storchi-Bergmann}, Thaisa},
        title = "{The Dust Content and Opacity of Actively Star-forming Galaxies}",
      journal = {\apj},
         year = 2000,
        month = apr,
       volume = {533},
       number = {2},
        pages = {682-695},
          doi = {10.1086/308692},
archivePrefix = {arXiv},
       eprint = {astro-ph/9911459},
 primaryClass = {astro-ph},
       adsurl = {https://ui.adsabs.harvard.edu/abs/2000ApJ...533..682C}
}

@ARTICLE{dale2014,
       author = {{Dale}, Daniel A. and {Helou}, George and {Magdis}, Georgios E. and {Armus}, Lee and {D{\'\i}az-Santos}, Tanio and {Shi}, Yong},
        title = "{A Two-parameter Model for the Infrared/Submillimeter/Radio Spectral Energy Distributions of Galaxies and Active Galactic Nuclei}",
      journal = {\apj},
         year = 2014,
        month = mar,
       volume = {784},
       number = {1},
          eid = {83},
        pages = {83},
          doi = {10.1088/0004-637X/784/1/83},
archivePrefix = {arXiv},
       eprint = {1402.1495},
 primaryClass = {astro-ph.GA},
       adsurl = {https://ui.adsabs.harvard.edu/abs/2014ApJ...784...83D}
}

@ARTICLE{Reines2015,
       author = {{Reines}, Amy E. and {Volonteri}, Marta},
        title = "{Relations between Central Black Hole Mass and Total Galaxy Stellar Mass in the Local Universe}",
      journal = {\apj},
         year = 2015,
        month = nov,
       volume = {813},
       number = {2},
          eid = {82},
        pages = {82},
          doi = {10.1088/0004-637X/813/2/82},
archivePrefix = {arXiv},
       eprint = {1508.06274},
 primaryClass = {astro-ph.GA},
       adsurl = {https://ui.adsabs.harvard.edu/abs/2015ApJ...813...82R}
}

@ARTICLE{Nicholl2018,
       author = {{Nicholl}, Matt},
        title = "{SuperBol: A User-friendly Python Routine for Bolometric Light Curves}",
      journal = {Research Notes of the American Astronomical Society},
         year = 2018,
        month = dec,
       volume = {2},
       number = {4},
          eid = {230},
        pages = {230},
          doi = {10.3847/2515-5172/aaf799},
       adsurl = {https://ui.adsabs.harvard.edu/abs/2018RNAAS...2..230N}
}

@ARTICLE{Lin2025,
       author = {{Lin}, Zheyu and {Jiang}, Ning and {Wang}, Yibo and {Kong}, Xu and {Huang}, Shifeng and {Lin}, Zesen and {Qin}, Chen and {Xia}, Tianyu},
        title = "{Insights from the ``Red Devil'' AT 2022fpx: A Dust-reddened Family of Tidal Disruption Events Excluded by Their Apparent Red Color?}",
      journal = {\apj},
         year = 2025,
        month = sep,
       volume = {990},
       number = {1},
          eid = {22},
        pages = {22},
          doi = {10.3847/1538-4357/adef10},
archivePrefix = {arXiv},
       eprint = {2507.04834},
 primaryClass = {astro-ph.HE},
       adsurl = {https://ui.adsabs.harvard.edu/abs/2025ApJ...990...22L}
}

@ARTICLE{Hung2018,
       author = {{Hung}, T. and {Gezari}, S. and {Cenko}, S.~B. and {van Velzen}, S. and {Blagorodnova}, N. and {Yan}, Lin and {Kulkarni}, S.~R. and {Lunnan}, R. and {Kupfer}, T. and {Leloudas}, G. and {Kong}, A.~K.~H. and {Nugent}, P.~E. and {Fremling}, C. and {Laher}, Russ R. and {Masci}, F.~J. and {Cao}, Y. and {Roy}, R. and {Petrushevska}, T.},
        title = "{Sifting for Sapphires: Systematic Selection of Tidal Disruption Events in iPTF}",
      journal = {\apjs},
         year = 2018,
        month = oct,
       volume = {238},
       number = {2},
          eid = {15},
        pages = {15},
          doi = {10.3847/1538-4365/aad8b1},
archivePrefix = {arXiv},
       eprint = {1712.04936},
 primaryClass = {astro-ph.HE},
       adsurl = {https://ui.adsabs.harvard.edu/abs/2018ApJS..238...15H}
}

@ARTICLE{Vv2019,
       author = {{van Velzen}, Sjoert and {Gezari}, Suvi and {Cenko}, S. Bradley and {Kara}, Erin and {Miller-Jones}, James C.~A. and {Hung}, Tiara and {Bright}, Joe and {Roth}, Nathaniel and {Blagorodnova}, Nadejda and {Huppenkothen}, Daniela and {Yan}, Lin and {Ofek}, Eran and {Sollerman}, Jesper and {Frederick}, Sara and {Ward}, Charlotte and {Graham}, Matthew J. and {Fender}, Rob and {Kasliwal}, Mansi M. and {Canella}, Chris and {Stein}, Robert and {Giomi}, Matteo and {Brinnel}, Valery and {van Santen}, Jakob and {Nordin}, Jakob and {Bellm}, Eric C. and {Dekany}, Richard and {Fremling}, Christoffer and {Golkhou}, V. Zach and {Kupfer}, Thomas and {Kulkarni}, Shrinivas R. and {Laher}, Russ R. and {Mahabal}, Ashish and {Masci}, Frank J. and {Miller}, Adam A. and {Neill}, James D. and {Riddle}, Reed and {Rigault}, Mickael and {Rusholme}, Ben and {Soumagnac}, Maayane T. and {Tachibana}, Yutaro},
        title = "{The First Tidal Disruption Flare in ZTF: From Photometric Selection to Multi-wavelength Characterization}",
      journal = {\apj},
         year = 2019,
        month = feb,
       volume = {872},
       number = {2},
          eid = {198},
        pages = {198},
          doi = {10.3847/1538-4357/aafe0c},
archivePrefix = {arXiv},
       eprint = {1809.02608},
 primaryClass = {astro-ph.HE},
       adsurl = {https://ui.adsabs.harvard.edu/abs/2019ApJ...872..198V}
}

@ARTICLE{Lin2018,
       author = {{Lin}, Dacheng and {Strader}, Jay and {Carrasco}, Eleazar R. and {Page}, Dany and {Romanowsky}, Aaron J. and {Homan}, Jeroen and {Irwin}, Jimmy A. and {Remillard}, Ronald A. and {Godet}, Olivier and {Webb}, Natalie A. and {Baumgardt}, Holger and {Wijnands}, Rudy and {Barret}, Didier and {Duc}, Pierre-Alain and {Brodie}, Jean P. and {Gwyn}, Stephen D.~J.},
        title = "{A luminous X-ray outburst from an intermediate-mass black hole in an off-centre star cluster}",
      journal = {Nature Astronomy},
         year = 2018,
        month = jun,
       volume = {2},
        pages = {656-661},
          doi = {10.1038/s41550-018-0493-1},
archivePrefix = {arXiv},
       eprint = {1806.05692},
 primaryClass = {astro-ph.HE},
       adsurl = {https://ui.adsabs.harvard.edu/abs/2018NatAs...2..656L}
}

@ARTICLE{Jin2025arXiv,
       author = {{Jin}, C.-C. and {Li}, D.-Y. and {Jiang}, N. and {Dai}, L.-X. and {Cheng}, H.-Q. and {Zhu}, J.-Z. and {Yang}, C.-W. and {Rau}, A. and {Baldini}, P. and {Wang}, T.-G. and {Zhou}, H.-Y. and {Yuan}, W. and {Zhang}, C. and {Shu}, X.-W. and {Shen}, R.-F. and {Wang}, Y.-L. and {Wen}, S.-X. and {Wu}, Q.-Y. and {Wang}, Y.-B. and {Thomsen}, L.~L. and {Zhang}, Z.-J. and {Zhang}, W.-J. and {Coleiro}, A. and {Eyles-Ferris}, R. and {Fang}, X. and {Ho}, L.~C. and {Hu}, J.-W. and {Jin}, J.-J. and {Li}, W.-X. and {Liu}, B.-F. and {Liu}, F.-K. and {Liu}, M.-J. and {Liu}, Z. and {Lu}, Y.-J. and {Merloni}, A. and {Qiao}, E.-L. and {Saxton}, R. and {Soria}, R. and {Wang}, S. and {Xue}, Y.-Q. and {Yang}, H.-N. and {Zhang}, B. and {Zhang}, W.-D. and {Cai}, Z.-M. and {Chen}, F.-S. and {Chen}, H.-L. and {Chen}, T.-X. and {Chen}, W. and {Chen}, Y.-H. and {Chen}, Y.-F. and {Chen}, Y. and {Cordier}, B. and {Cui}, C.-Z. and {Cui}, W.-W. and {Dai}, Y.-F. and {Ding}, H.-C. and {Fan}, D.-W. and {Fan}, Z. and {Feng}, H. and {Garcia}, J.~A. and {Guan}, J. and {Han}, D.-W. and {Hou}, D.-J. and {Hu}, H.-B. and {Huang}, M.-H. and {Huo}, J. and {Jia}, S.-M. and {Jia}, Z.-Q. and {Jiang}, B.-W. and {Jin}, G. and {Kong}, X. and {Kuulkers}, E. and {Lei}, W.-H. and {Li}, C.-K. and {Li}, J.-F. and {Li}, L.-H. and {Li}, M.-S. and {Li}, W. and {Li}, Z.-D. and {Lian}, T.-Y. and {Ling}, Z.-X. and {Liu}, C.-Z. and {Liu}, H.-Y and {Liu}, H.-Q. and {Liu}, J.-F. and {Liu}, Y. and {Lu}, F.-J. and {Luo}, L.-D. and {Ma}, J. and {Mao}, X. and {Mu}, H.-Y. and {Nandra}, K. and {O'Brien}, P. and {Pan}, H.-W. and {Pan}, X. and {Qin}, G.-J. and {Rea}, N. and {Sanders}, J. and {Song}, L.-M. and {Sun}, H. and {Sun}, S.-L. and {Sun}, X.-J. and {Tan}, Y.-Y. and {Tang}, Q.-J. and {Tao}, Y.-H. and {Wang}, B.-C. and {Wang}, J. and {Wang}, J.-F. and {Wang}, L. and {Wang}, W.-X. and {Wang}, Y.-S. and {Wang}, Z.-X. and {Wu}, Q.-W. and {Wu}, X.-F. and {Xu}, H.-T. and {Xu}, J.-J. and {Xu}, X.-P. and {Xu}, Y.-F. and {Xu}, Z. and {Xue}, C.-B. and {Xue}, S.-J. and {Xue}, Y.-L. and {Yan}, A.-L. and {Yang}, X.-T. and {Yang}, Y.-J. and {Zhang}, J. and {Zhang}, M. and {Zhang}, S.-N. and {Zhang}, Y.-H. and {Zhang}, Z. and {Zhang}, Z. and {Zhang}, Z.-L. and {Zhao}, D.-H. and {Zhao}, H.-S. and {Zhao}, X.-F. and {Zhao}, Z.-J. and {Zheng}, J. and {Zhu}, Q.-F. and {Zhu}, Y.-X. and {Zhu}, Z.-C. and {Zou}, H.},
        title = "{An Intermediate-mass Black Hole Lurking in A Galactic Halo Caught Alive during Outburst}",
      journal = {arXiv e-prints},
         year = 2025,
        month = jan,
          eid = {arXiv:2501.09580},
        pages = {arXiv:2501.09580},
          doi = {10.48550/arXiv.2501.09580},
archivePrefix = {arXiv},
       eprint = {2501.09580},
 primaryClass = {astro-ph.HE},
       adsurl = {https://ui.adsabs.harvard.edu/abs/2025arXiv250109580J}
}

@ARTICLE{Yao2025,
       author = {{Yao}, Yuhan and {Chornock}, Ryan and {Ward}, Charlotte and {Hammerstein}, Erica and {Sfaradi}, Itai and {Margutti}, Raffaella and {Kelley}, Luke Zoltan and {Lu}, Wenbin and {Liu}, Chang and {Wise}, Jacob and {Sollerman}, Jesper and {Alexander}, Kate D. and {Bellm}, Eric C. and {Drake}, Andrew J. and {Fremling}, Christoffer and {Gilfanov}, Marat and {Graham}, Matthew J. and {Groom}, Steven L. and {Hinds}, K.~R. and {Kulkarni}, S.~R. and {Miller}, Adam A. and {Miller-Jones}, James C.~A. and {Nicholl}, Matt and {Perley}, Daniel A. and {Purdum}, Josiah and {Ravi}, Vikram and {Rich}, R. Michael and {Rehemtulla}, Nabeel and {Riddle}, Reed and {Smith}, Roger and {Stein}, Robert and {Sunyaev}, Rashid and {van Velzen}, Sjoert and {Wold}, Avery},
        title = "{A Massive Black Hole 0.8 kpc from the Host Nucleus Revealed by the Offset Tidal Disruption Event AT2024tvd}",
      journal = {\apjl},
         year = 2025,
        month = jun,
       volume = {985},
       number = {2},
          eid = {L48},
        pages = {L48},
          doi = {10.3847/2041-8213/add7de},
archivePrefix = {arXiv},
       eprint = {2502.17661},
 primaryClass = {astro-ph.GA},
       adsurl = {https://ui.adsabs.harvard.edu/abs/2025ApJ...985L..48Y}
}

@ARTICLE{Stein2026,
       author = {{Stein}, Robert and {Carney}, Jonathan and {Ward}, Charlotte and {Margutti}, Raffaella and {Hall}, Xander J. and {Sfaradi}, Itai and {Andreoni}, Igor and {Chornock}, Ryan and {Gezari}, Suvi and {Mo}, Geoffrey and {Yao}, Yuhan and {Bellm}, Eric C. and {Bloom}, Joshua S. and {Busmann}, Malte and {Caiazzo}, Ilaria and {Cenko}, S. Bradley and {Graham}, Matthew J. and {Groom}, Steven L. and {Gruen}, Daniel and {Hammerstein}, Erica and {Kasliwal}, Mansi M. and {O'Connor}, Brendan and {Palmese}, Antonella and {Purdum}, Josiah and {Rastinejad}, Jillian C. and {Riddle}, Reed and {Rusholme}, Ben and {Sollerman}, Jesper and {Somalwar}, Jean J. and {Veilleux}, Sylvain},
        title = "{TDE 2025abcr: A Tidal Disruption Event in the Outskirts of a Massive Galaxy}",
      journal = {arXiv e-prints},
         year = 2026,
        month = feb,
          eid = {arXiv:2602.10180},
        pages = {arXiv:2602.10180},
          doi = {10.48550/arXiv.2602.10180},
archivePrefix = {arXiv},
       eprint = {2602.10180},
 primaryClass = {astro-ph.HE},
       adsurl = {https://ui.adsabs.harvard.edu/abs/2026arXiv260210180S}
}

@ARTICLE{Vv2020SSRv,
       author = {{van Velzen}, Sjoert and {Holoien}, Thomas W.-S. and {Onori}, Francesca and {Hung}, Tiara and {Arcavi}, Iair},
        title = "{Optical-Ultraviolet Tidal Disruption Events}",
      journal = {\ssr},
         year = 2020,
        month = oct,
       volume = {216},
       number = {8},
          eid = {124},
        pages = {124},
          doi = {10.1007/s11214-020-00753-z},
archivePrefix = {arXiv},
       eprint = {2008.05461},
 primaryClass = {astro-ph.HE},
       adsurl = {https://ui.adsabs.harvard.edu/abs/2020SSRv..216..124V}
}

@ARTICLE{Ivezi2019,
       author = {{Ivezi{\'c}}, {\v{Z}}eljko and {Kahn}, Steven M. and {Tyson}, J. Anthony and {Abel}, Bob and {Acosta}, Emily and {Allsman}, Robyn and {Alonso}, David and {AlSayyad}, Yusra and {Anderson}, Scott F. and {Andrew}, John and {Angel}, James Roger P. and {Angeli}, George Z. and {Ansari}, Reza and {Antilogus}, Pierre and {Araujo}, Constanza and {Armstrong}, Robert and {Arndt}, Kirk T. and {Astier}, Pierre and {Aubourg}, {\'E}ric and {Auza}, Nicole and {Axelrod}, Tim S. and {Bard}, Deborah J. and {Barr}, Jeff D. and {Barrau}, Aurelian and {Bartlett}, James G. and {Bauer}, Amanda E. and {Bauman}, Brian J. and {Baumont}, Sylvain and {Bechtol}, Ellen and {Bechtol}, Keith and {Becker}, Andrew C. and {Becla}, Jacek and {Beldica}, Cristina and {Bellavia}, Steve and {Bianco}, Federica B. and {Biswas}, Rahul and {Blanc}, Guillaume and {Blazek}, Jonathan and {Blandford}, Roger D. and {Bloom}, Josh S. and {Bogart}, Joanne and {Bond}, Tim W. and {Booth}, Michael T. and {Borgland}, Anders W. and {Borne}, Kirk and {Bosch}, James F. and {Boutigny}, Dominique and {Brackett}, Craig A. and {Bradshaw}, Andrew and {Brandt}, William Nielsen and {Brown}, Michael E. and {Bullock}, James S. and {Burchat}, Patricia and {Burke}, David L. and {Cagnoli}, Gianpietro and {Calabrese}, Daniel and {Callahan}, Shawn and {Callen}, Alice L. and {Carlin}, Jeffrey L. and {Carlson}, Erin L. and {Chandrasekharan}, Srinivasan and {Charles-Emerson}, Glenaver and {Chesley}, Steve and {Cheu}, Elliott C. and {Chiang}, Hsin-Fang and {Chiang}, James and {Chirino}, Carol and {Chow}, Derek and {Ciardi}, David R. and {Claver}, Charles F. and {Cohen-Tanugi}, Johann and {Cockrum}, Joseph J. and {Coles}, Rebecca and {Connolly}, Andrew J. and {Cook}, Kem H. and {Cooray}, Asantha and {Covey}, Kevin R. and {Cribbs}, Chris and {Cui}, Wei and {Cutri}, Roc and {Daly}, Philip N. and {Daniel}, Scott F. and {Daruich}, Felipe and {Daubard}, Guillaume and {Daues}, Greg and {Dawson}, William and {Delgado}, Francisco and {Dellapenna}, Alfred and {de Peyster}, Robert and {de Val-Borro}, Miguel and {Digel}, Seth W. and {Doherty}, Peter and {Dubois}, Richard and {Dubois-Felsmann}, Gregory P. and {Durech}, Josef and {Economou}, Frossie and {Eifler}, Tim and {Eracleous}, Michael and {Emmons}, Benjamin L. and {Fausti Neto}, Angelo and {Ferguson}, Henry and {Figueroa}, Enrique and {Fisher-Levine}, Merlin and {Focke}, Warren and {Foss}, Michael D. and {Frank}, James and {Freemon}, Michael D. and {Gangler}, Emmanuel and {Gawiser}, Eric and {Geary}, John C. and {Gee}, Perry and {Geha}, Marla and {Gessner}, Charles J.~B. and {Gibson}, Robert R. and {Gilmore}, D. Kirk and {Glanzman}, Thomas and {Glick}, William and {Goldina}, Tatiana and {Goldstein}, Daniel A. and {Goodenow}, Iain and {Graham}, Melissa L. and {Gressler}, William J. and {Gris}, Philippe and {Guy}, Leanne P. and {Guyonnet}, Augustin and {Haller}, Gunther and {Harris}, Ron and {Hascall}, Patrick A. and {Haupt}, Justine and {Hernandez}, Fabio and {Herrmann}, Sven and {Hileman}, Edward and {Hoblitt}, Joshua and {Hodgson}, John A. and {Hogan}, Craig and {Howard}, James D. and {Huang}, Dajun and {Huffer}, Michael E. and {Ingraham}, Patrick and {Innes}, Walter R. and {Jacoby}, Suzanne H. and {Jain}, Bhuvnesh and {Jammes}, Fabrice and {Jee}, M. James and {Jenness}, Tim and {Jernigan}, Garrett and {Jevremovi{\'c}}, Darko and {Johns}, Kenneth and {Johnson}, Anthony S. and {Johnson}, Margaret W.~G. and {Jones}, R. Lynne and {Juramy-Gilles}, Claire and {Juri{\'c}}, Mario and {Kalirai}, Jason S. and {Kallivayalil}, Nitya J. and {Kalmbach}, Bryce and {Kantor}, Jeffrey P. and {Karst}, Pierre and {Kasliwal}, Mansi M. and {Kelly}, Heather and {Kessler}, Richard and {Kinnison}, Veronica and {Kirkby}, David and {Knox}, Lloyd and {Kotov}, Ivan V. and {Krabbendam}, Victor L. and {Krughoff}, K. Simon and {Kub{\'a}nek}, Petr and {Kuczewski}, John and {Kulkarni}, Shri and {Ku}, John and {Kurita}, Nadine R. and {Lage}, Craig S. and {Lambert}, Ron and {Lange}, Travis and {Langton}, J. Brian and {Le Guillou}, Laurent and {Levine}, Deborah and {Liang}, Ming and {Lim}, Kian-Tat and {Lintott}, Chris J. and {Long}, Kevin E. and {Lopez}, Margaux and {Lotz}, Paul J. and {Lupton}, Robert H. and {Lust}, Nate B. and {MacArthur}, Lauren A. and {Mahabal}, Ashish and {Mandelbaum}, Rachel and {Markiewicz}, Thomas W. and {Marsh}, Darren S. and {Marshall}, Philip J. and {Marshall}, Stuart and {May}, Morgan and {McKercher}, Robert and {McQueen}, Michelle and {Meyers}, Joshua and {Migliore}, Myriam and {Miller}, Michelle and {Mills}, David J.},
        title = "{LSST: From Science Drivers to Reference Design and Anticipated Data Products}",
      journal = {\apj},
         year = 2019,
        month = mar,
       volume = {873},
       number = {2},
          eid = {111},
        pages = {111},
          doi = {10.3847/1538-4357/ab042c},
archivePrefix = {arXiv},
       eprint = {0805.2366},
 primaryClass = {astro-ph},
       adsurl = {https://ui.adsabs.harvard.edu/abs/2019ApJ...873..111I}
}

@ARTICLE{Jiang2021,
       author = {{Jiang}, Ning and {Wang}, Tinggui and {Dou}, Liming and {Shu}, Xinwen and {Hu}, Xueyang and {Liu}, Hui and {Wang}, Yibo and {Yan}, Lin and {Sheng}, Zhenfeng and {Yang}, Chenwei and {Sun}, Luming and {Zhou}, Hongyan},
        title = "{Mid-infrared Outbursts in Nearby Galaxies (MIRONG). I. Sample Selection and Characterization}",
      journal = {\apjs},
         year = 2021,
        month = feb,
       volume = {252},
       number = {2},
          eid = {32},
        pages = {32},
          doi = {10.3847/1538-4365/abd1dc},
archivePrefix = {arXiv},
       eprint = {2012.06806},
 primaryClass = {astro-ph.GA},
       adsurl = {https://ui.adsabs.harvard.edu/abs/2021ApJS..252...32J}
}

@ARTICLE{Zhu2023,
       author = {{Zhu}, Jiazheng and {Jiang}, Ning and {Wang}, Tinggui and {Huang}, Shifeng and {Lin}, Zheyu and {Wang}, Yibo and {Wang}, Jian-Guo},
        title = "{AT 2023clx: The Faintest and Closest Optical Tidal Disruption Event Discovered in Nearby Star-forming Galaxy NGC 3799}",
      journal = {\apjl},
         year = 2023,
        month = aug,
       volume = {952},
       number = {2},
          eid = {L35},
        pages = {L35},
          doi = {10.3847/2041-8213/ace625},
archivePrefix = {arXiv},
       eprint = {2307.04297},
 primaryClass = {astro-ph.HE},
       adsurl = {https://ui.adsabs.harvard.edu/abs/2023ApJ...952L..35Z}
}

@ARTICLE{Li2026,
       author = {{Li}, Wenkai and {Christy}, Collin T. and {Alexander}, Kate D. and {Sfaradi}, Itai and {Huang}, Xinya and {Jiang}, Ning and {Laskar}, Tanmoy and {Mummery}, Andrew and {Franz}, Noah and {Goodwin}, Adelle J. and {Golay}, Walter W. and {Margutti}, Raffaella and {Chornock}, Ryan and {Zhu}, Jiazheng and {van Velzen}, Sjoert and {Cendes}, Yvette and {Lu}, Wenbin and {Lynch}, Jimmy},
        title = "{VLA Observations Confirm AT 2023mfm as an Off-nuclear Tidal Disruption Event}",
      journal = {Research Notes of the American Astronomical Society},
         year = 2026,
        month = jun,
       volume = {10},
       number = {6},
          eid = {153},
        pages = {153},
          doi = {10.3847/2515-5172/ae7a6a},
       adsurl = {https://ui.adsabs.harvard.edu/abs/2026RNAAS..10..153L}
}

@ARTICLE{Zhu2026,
       author = {{Zhu}, Jiazheng and {Xu}, Zelin and {Jiang}, Ning and {Jiang}, Ji-an and {Wang}, Tinggui and {Yao}, Yuhan and {Chornock}, Ryan and {Hammerstein}, Erica and {Wang}, Yibo and {Cai}, Min-Xuan and {Huang}, Shifeng and {Li}, Wenkai and {Wu}, Mingxin and {Jin}, Chichuan and {Lin}, Jie and {Lyu}, Jianwei and {Meng}, Dezheng and {Wu}, Weiyu and {Liu}, Zhengyan and {Zhao}, Junhan and {Jia}, Ziqing and {Wang}, Chengyi and {Fan}, Lulu and {Kong}, Xu and {Li}, Feng and {Liang}, Ming and {Tang}, Jinling and {Wang}, Hairen and {Wang}, Jian and {Xue}, Yongquan and {Yang}, Ji and {Zhang}, Hongfei and {Zhao}, Wen and {Zhu}, Qingfeng},
        title = "{Discovery of a Featureless Tidal Disruption Event at z {\ensuremath{\sim}} 1 with the Wide Field Survey Telescope}",
      journal = {\apjl},
         year = 2026,
        month = jun,
       volume = {1004},
       number = {1},
          eid = {L10},
        pages = {L10},
          doi = {10.3847/2041-8213/ae710d},
archivePrefix = {arXiv},
       eprint = {2605.21950},
 primaryClass = {astro-ph.HE},
       adsurl = {https://ui.adsabs.harvard.edu/abs/2026ApJ..1004L..10Z}
}

@ARTICLE{cyj,
       author = {{Hammerstein}, E. and {Chornock}, R. and {Stein}, R. and {Somalwar}, J. and {Carney}, J. and {Andreoni}, I. and {Yao}, Y. and {LeBaron}, N. and {Guo}, X. and {Ravi}, V.},
        title = "{ZTF Classification of ZTF25aagfhmz (AT2025cyj), ZTF25aaknwwy (AT2025hee), and ZTF25aagrecw (AT2025gmz) as TDEs}",
      journal = {Transient Name Server AstroNote},
         year = 2025,
        month = apr,
       volume = {127},
        pages = {1},
       adsurl = {https://ui.adsabs.harvard.edu/abs/2025TNSAN.127....1H}
}

@ARTICLE{aapf,
       author = {{Chornock}, R. and {Stein}, R. and {Hammerstein}, E. and {Carney}, J.},
        title = "{ZTF Classification of AT 2025aapf/ZTF25abxcnsf as a likely TDE-featureless}",
      journal = {Transient Name Server AstroNote},
         year = 2025,
        month = oct,
       volume = {309},
        pages = {1},
       adsurl = {https://ui.adsabs.harvard.edu/abs/2025TNSAN.309....1C}
}

@ARTICLE{afvr1,
       author = {{Charalampopoulos}, P.},
        title = "{Tentative classification of the optical transient AT 2025afvr as a featureless tidal disruption event.}",
      journal = {Transient Name Server AstroNote},
         year = 2025,
        month = dec,
       volume = {340},
        pages = {1},
       adsurl = {https://ui.adsabs.harvard.edu/abs/2025TNSAN.340....1C}
}

@ARTICLE{afvr2,
       author = {{Wu}, M. and {Wang}, Y. and {Huang}, S. and {Xu}, L. and {Jiang}, N. and {Wang}, T.},
        title = "{Optical spectroscopic and Swift observations of TDE 2025afvr}",
      journal = {Transient Name Server AstroNote},
         year = 2025,
        month = dec,
       volume = {342},
        pages = {1},
       adsurl = {https://ui.adsabs.harvard.edu/abs/2025TNSAN.342....1W}
}

@ARTICLE{afvr3,
       author = {{Hammerstein}, E. and {Chornock}, R. and {Carney}, J. and {Stein}, R. and {Somalwar}, J.},
        title = "{ZTF classification of two new TDEs and one Bowen Fluorescence Flare, and spectra/redshifts for three additional TDEs}",
      journal = {Transient Name Server AstroNote},
         year = 2026,
        month = feb,
       volume = {37},
        pages = {1},
       adsurl = {https://ui.adsabs.harvard.edu/abs/2026TNSAN..37....1H}
}

@ARTICLE{aljd,
       author = {{Chornock}, R. and {Hammerstein}, E. and {Stein}, R. and {Carney}, J.},
        title = "{ZTF Classification of AT 2025aljd/ZTF25aclckzl as a likely TDE}",
      journal = {Transient Name Server AstroNote},
         year = 2026,
        month = jan,
       volume = {7},
        pages = {1},
       adsurl = {https://ui.adsabs.harvard.edu/abs/2026TNSAN...7....1C}
}

@ARTICLE{dta,
       author = {{Hammerstein}, E. and {Chornock}, R. and {Stein}, R.},
        title = "{ZTF classification of ZTF26aalndlm / AT2026hdo and ZTF26aahjwyr / AT2026dta as tidal disruption events}",
      journal = {Transient Name Server AstroNote},
         year = 2026,
        month = mar,
       volume = {85},
        pages = {1},
       adsurl = {https://ui.adsabs.harvard.edu/abs/2026TNSAN..85....1H}
}

\end{document}